\documentclass[11pt,a4paper]{article}

\usepackage{graphicx}

\usepackage{cite}

\usepackage{hyperref}

\usepackage{geometry}
\begin{document}
\title{Nonsingular Bouncing Model in Closed and Open universe\\ } 

\date{\today}

\author{Manabendra Sharma\footnote{email: manabendra@iiserb.ac.in}}

\maketitle

\centerline{Department of Physics, IISER Bhopal, Bhopal - 462023, India}

\begin{abstract} 
  We investigate the cosmology of a class of model with noncanonical scalar field and matter both in FRW closed and open background. Writing the Einstein Equations in terms of dimensionless dynamical variables suitable for studying bouncing solutions, a fixed point analysis is carried out. Cosmological solutions satisfying the stability and bouncing conditions are obtained.

\end{abstract}

\section{Introduction}
There are two scenarios exist in literature, namely, inflation and bouncing model which address the shortcomings of the standard model of cosmology.
Though inflation solves most of the problems (horizon, flatness and entropy)
of the standard model of cosmology, the issue with the initial singularity is not resolved under its domain. It is the alternate scenario, nonsingular bouncing model, that eradicates the singularity by constructing a universe which begins with a contracting phase and then bounces back to an expanding phase through a non zero minimum in the scale factor. Nonsingular bouncing models can be categorised into two types, matter bounce model\cite{Fabio} and and Ekpyrotik models\cite{khoury1}. For a review on these models refer to \cite{RHBrev} and \cite{Lehners}. 


In this paper we consider a noncanonical scalar field with a general function of kinetic term $F(X),$ where $X= - 1/2 \partial_{\mu}\phi \partial^{\mu}\phi.$ These theories are originally motivated to provide a large tensor to scalar perturbation in inflationary settings\cite{Mukhanov:2005bu, Panotopoulos:2007ky, Unnikrishnan:2012zu}. Dark energy with a general kinetic term $F(X)$ is modeled first in ref. \cite{Chiba:1999ka}. For other variants of models of dark energy in this context refer to \cite{Copeland:2006wr}. Other works related to unifying dark matter, dark energy and$/$or inflation for noncanonical scalar field models are studied in \cite{Bertacca:2010ct, Bose:2008ew, Bose:2009kc, DeSantiago:2011qb}. In order to study the phase space in this model, we write the first order equations of motion in terms of dimensionless dynamical variables \cite{Copeland:1997et}. The motivation to use noncanonical scalar field as matter is to construct nonsingular bouncing models. The phase space analysis of a cosmological model with scalar field Lagrangian $F(X)-V(\phi)$ and matter for an FRW background is given in ref.\cite{wands}. The condition for nonsingular bounce is also discussed in ref.\cite{wands}. In order to explore the behaviour of curvature parameter near bounce in an nonsingular bouncing model we do a phase space analysis in an FRW closed and open universe. This can be easily extended to other nonsingular bouncing models. 

We study the cosmology of an anisotropic universe with a matter Lagrangian of the form $F(X)-V(\phi)$ and matter. In section \ref{EFE} we write the Einstein's equation in terms of dynamical variables suitable for the analysis of a bouncing scenario in FRW closed and open universe. Then we discuss fixed points and their stability in section \ref{FPA}. Conditions for existence of nonsingular bouncing solution is derived in terms of dynamical variables in section \ref{BouncingScenario}. We summarise our results in section \ref{Conclusion}.

\section{Einstein Equations in Bianchi I background}\label{EFE}
The action for our model is given by
\begin{equation}
S= \int d^4 x \sqrt{-g} [\frac{1}{2} R + F(X) - V(\phi) + L_m] 
\label{}
\end{equation}
where $L_m$ is the lagrangian of the matter field.

 To see the behaviour of curvature parameter of the spacetime in a nonsingular bouncing scenario we work with a FRW closed and open universe. The line element of the same is given by:

\begin{equation}
ds^2=-dt^2+a^2(t)[\frac{dr^2}{1-kr^2}+r^2d\theta ^{2}+r^{2}sin^{2}\theta {d\phi}^2].
\end{equation}

where $k=+1$ denotes closed and $k=-1$ denotes an open universe respectively.



The Hubble parameter $H$ is defined as 

\begin{equation}
H = {\frac{1}{a}}{\frac{\mathrm{d} a}{\mathrm{d} t}}  \nonumber\\
\end{equation}

In terms of Hubble parameter, the Einstein equations take the following form 
\begin{eqnarray}
\frac{\mathrm{d} H}{\mathrm{d} t} & = &-H^2-\frac{1}{6}(\rho+3p) \,, \nonumber\\
H^2 &=& \frac{\rho}{3}-\frac{k}{a^2} \,,
\label{einstein} 
\end{eqnarray}

where $\rho=\rho_{\phi}+\rho_m$ and $p=p_{\phi}+p_m.$

Here the energy density $\rho_{\phi}$ and pressure $p_{\phi}$ of the scalar field is found to be


\begin{eqnarray}
\rho_{\phi}&=&2XF_{X}-F+V \,, \nonumber\\
  p_{\phi}&=&F(X)-V(\phi) \,,
\label{scalarenergydensity}
\end{eqnarray}

and $\rho_m$ and $p_m$ are the energy density and pressure due to the term $L_m$. 
 
Substituting Eq.(\ref{scalarenergydensity}) in first and third line of Eq.(\ref{einstein}), we get

\begin{equation}
\frac{\mathrm{d} H}{\mathrm{d} t}=-H^2-\frac{1}{6}(2XF_X-F+V+\rho_m+3(F-V)+3p_m)
\label{rc}
\end{equation}

\begin{equation}
H^2=\frac{2XF_X-F}{3}+\frac{V}{3}-\frac{k}{a^2}+\frac{\rho_m}{3}
\end{equation}

Here we further define few more variables which are useful for defining dimensionless dynamical variables. They are
\begin{eqnarray}
\rho_k&=&2XF_{X}-F \,, \nonumber\\
w_k&=&\frac{F}{2XF_{X}-F} \,, \nonumber\\
\sigma&=&-\frac{1}{\sqrt3 |\rho_k|}\frac{\mathrm{d} logV}{\mathrm{d} t} \,,
\label{parameters} 
\end{eqnarray}
where $\rho_k$ is the kinetic part of the energy density $\rho_{\phi},$  $w_k$ is the ratio of kinetic part of the pressure $p_{\phi}$ to the $\rho_k$ and $\sigma$ is the auxiliary variable which depends on the variation of potential with time.  

Neglecting the interaction between scalar field and  matter, continuity equation for $\rho_{\phi}$ in terms of dimensionless time variable N $(dN= H dt),$ is
\begin{equation}
\frac{\mathrm{d}}{\mathrm{d} N}(2XF_X-F+V)+6XF_X=0. 
\label{continuity}
\end{equation}

 Now we define a set of  dimensionless dynamical variables which is suitable for nonsingular bounce models. Relevance of these variables that they remain finite during the entire evolution across bounce. The dynamical variables are

\begin{eqnarray}
\tilde{x}=\frac{\sqrt{3}H }{\sqrt{|\rho_k|}} \,, 
\tilde{y}=\frac{\sqrt{|V|} }{\sqrt{|\rho_k|}}sign(V) \,, 
\tilde{z}=\frac{\sqrt{|k|}}{\sqrt{|\rho_k}}sign(\tilde z) \,,
\tilde{\Omega}_m=\frac{\rho_m}{|\rho_k|} .
\label{bouncingvariables1}
\end{eqnarray}

Here $sign(\tilde z)\equiv$$sign(k)$ denotes FRW closed universe for $+1$ and open for $-1.$ Using Eqs.(\ref{bouncingvariables1}), (\ref{einstein}) and (\ref{continuity}) and parameters defined in Eq.(\ref{parameters}), the evolution equations of $\tilde{x},$ $\tilde{y}$ and $\tilde{z}$ are written as,

\begin{eqnarray}
\frac{\mathrm{d} \tilde{x}}{\mathrm{d} \tilde{N}}&=&-\frac{3}{2}\left [ (w_k-w_m)sign(\rho_k)+(1+w_m)(\tilde{x}^2-\tilde{y}\tilde{|y|})+\frac{\tilde{z}|\tilde{z}
|}{3}(1+3w_m) \right ]\nonumber\\&+&\frac{3}{2}\tilde{x}\left [(w_k+1)\tilde{x}-\sigma \tilde{y}|\tilde{y}|sign(\rho_k)  \right ] , \nonumber\\
\frac{\mathrm{d} \tilde{y}}{\mathrm{d} \tilde{N}}&=&\frac{3}{2}\tilde{y}\left [ -\sigma+(w_k+1) \tilde{x}-\sigma \tilde{y}{|\tilde{y}|}sign(\rho_k)\right ] , \nonumber\\
\frac{\mathrm{d} \tilde{z}}{\mathrm{d} \tilde{N}}&=&-\tilde{z}\tilde{x}+\frac{3}{2}\tilde{z}(\tilde{x}(1+w_k)-\tilde{y}|\tilde{y}|sign(\rho_k)) , \nonumber\\
\frac{\mathrm{d} \tilde{\Omega}_m}{\mathrm{d} \tilde{N}}&=&-3(1+w_m)\tilde{x}\tilde{\Omega}_m-\tilde{\Omega}_m\left [ 3\sigma \tilde{y}|\tilde{y}|sign(\rho_k)-3\tilde{x}(1+w_k) \right ] \,,  \nonumber\\
\end{eqnarray}

where $d \tilde{N}=\sqrt{\frac{|\rho_k|}{3}} dt$
and the constraint equation relating dynamical variables is
\begin{equation}
\tilde{x}^2-\tilde{y}|\tilde{y}|+{\tilde z|\tilde z|}-\tilde{\Omega}_m=1\times sign(\rho_k) .
\label{constraint}
\end{equation}

The equation for parameter $\sigma$ becomes \cite{wands}
\begin{equation}
\frac{\mathrm{d} \sigma}{\mathrm{d} \tilde N}= -3\sigma^2\left ( \Gamma-1\right )+\frac{3\sigma\left ( 2\Xi \left ( w_k+1 \right )+w_k-1 \right )}{2\left ( 2\sigma +1 \right )\left ( w_k+1 \right )}\left [ \left ( w_k+1 \right )\tilde x-\sigma \tilde{y}^2 \right ]\, 
\end{equation}
where $\Xi=\frac{X F_{XX}}{F_X}$ and $\Gamma = \frac{V V_{\phi \phi}}{V_{\phi}}.$   
 
For our model we have taken power law form for $F(X)=F_0 X^{\eta},$ where $F_0$ is a constant. For this form of $F(X),$  $w_k=\frac{1}{2 \eta -1}$ and $\Xi=\eta -1.$

Potential $V(\phi)$ is taken as $V(\phi)=V_0e^{-c \phi},$ where $V_0$ and c are constants with positive values. For this choice of $V(\phi),$  $ \Gamma$ becomes unity.   

In the next section, we do a fixed point analysis of dynamical equations for $\tilde x$, $\tilde y$, $\tilde z$ and $\sigma.$ The evolution of $\tilde{\Omega}_ m$ is determined from the constraint Eq.(\ref{constraint}).

\section{Fixed Point Analysis}\label{FPA}

 In this section, we do a fixed point analysis of our system of dynamical equation in order to extract the qualitative information about the nature of solution. Fixed points are calculated by taking the first derivative of the dynamical variables to be zero. The stability of a fixed point is determined from the behaviour of a small perturbation around that  fixed point. 

 We get the set of fixed points $\tilde{x}_c$, $\tilde{y}_c$, $\tilde{z}_c$ and ${\sigma}_c$ by solving the following set of equations simultaneously (where the subscript c denotes fixed points). Now, if we define the slopes of the dynamical variables $\tilde{x}$, $\tilde{y}$, $\tilde{z}$ and ${\sigma}$ as $f(\tilde{x},\tilde{y},\tilde{z},\sigma)$, $g(\tilde{x},\tilde{y},\tilde{z},\sigma)$, $h(\tilde{x},\tilde{y},\tilde{z},\sigma)$ and $i(\tilde{x},\tilde{y},\tilde{z},\sigma)$. The set of equations we need to solve to obtain the fixed point is

\begin{eqnarray}
f(\tilde{x},\tilde{y},\tilde{z},{\sigma}) &\equiv& \frac{\mathrm{d} \tilde{x}}{\mathrm{d} \tilde{N}}=0 \,,  \nonumber \\
g(\tilde{x},\tilde{y},\tilde{z},{\sigma}) &\equiv& \frac{\mathrm{d} \tilde{y}}{\mathrm{d} \tilde{N}}=0 \,, \nonumber \\
h(\tilde{x},\tilde{y},\tilde{z},{\sigma}) &\equiv& \frac{\mathrm{d} \tilde{z}}{\mathrm{d} \tilde{N}}=0 \,, \nonumber \\
i(\tilde{x},\tilde{y},\tilde{z},{\sigma}) &\equiv& \frac{\mathrm{d} {\sigma}}{\mathrm{d} \tilde{N}}=0 \,, \nonumber \\
\end{eqnarray} 

where,
\begin{eqnarray}
f(\tilde x,\tilde y,\tilde z,\sigma) &\equiv& -\frac{3}{2}[(w_k-w_m)(sign\rho_k)+(1+w_m)(\tilde{x}^2-\tilde{y}|\tilde{y}|)+(1+3w_m)\frac{\tilde{z}|\tilde z|}{3}]\nonumber\\&+&\frac{3}{2}\tilde{x}[(w_k+1)\tilde{x}-\sigma \tilde{y}|\tilde{y}|sign(\rho_k)]\,, \nonumber \\
g(\tilde x,\tilde y,\tilde z,\sigma) &\equiv& \frac{3}{2}\tilde{y}[-\sigma+(w_k+1)\tilde{x}-\sigma\tilde{y}|\tilde{y}|sign(\rho_k)]\,, \nonumber \\
h(\tilde x,\tilde y,\tilde z,\sigma) &\equiv& -3\tilde{z}\tilde{x}+3\tilde{z}\tilde{x}(1+w_k)-3\tilde{z}\tilde{y}|\tilde{y}|sign(\rho_k)\,, \nonumber \\
i(\tilde x,\tilde y,\tilde z,\sigma) &\equiv& \frac{3}{2}\frac{[2\Xi (w_k+1)+(w_k-1)]}{2(2\sigma+1)(w_k+1)}[(w_k+1)\tilde{x}-\sigma \tilde{y}^2].
\end{eqnarray}

The corresponding fixed point for $\tilde{\Omega}_m$ can be found using the constraint Eq.(\ref{constraint}).

The stability of the fixed points can be examined from the evolution of pertubations around fixed points. 
Now, if $(\tilde{x}_c, \tilde{y}_c, \tilde{z}_c, {\sigma}_c)$ is a fixed point and $\delta \tilde{x}= \tilde{x}-\tilde{x}_c, \delta \tilde{y}=\tilde{y}-\tilde{y}_c$,
 $\delta \tilde z= \tilde{z}-\tilde{z}_c$  and $\delta \sigma= \sigma-\sigma_c$ be the respective perturbation around it, then the evolution of the perturbation is determined by

\begin{eqnarray}
   \delta \dot{\tilde {x}} &=& \dot{\tilde x}= f(\tilde{x}_c+\delta \tilde {x},\tilde {y}_c+\delta \tilde {y},\tilde {z}_c+\delta \tilde {z},  {\sigma} +\delta {\sigma} ) \,, \nonumber \\
   \delta \dot{\tilde {y}}&=& \dot{\tilde y}= g(\tilde{x}_c+\delta \tilde {x},\tilde {y}_c+\delta \tilde {y},\tilde {z}_c+\delta \tilde {z}, {\sigma}+\delta {\sigma}) \,, \nonumber \\
  \delta \dot{\tilde {z}}&=& \dot{\tilde z}= h(\tilde{x}_c+\delta \tilde {x},\tilde {y}_c+\delta \tilde {y},\tilde {z}_c+\delta \tilde {z}, {\sigma}+\delta {\sigma})\,, \nonumber\\  
  \delta \dot{ {\sigma}}&=& \dot{\sigma}= h(\tilde{x}_c+\delta \tilde {x},\tilde {y}_c+\delta \tilde {y},\tilde {z}_c+\delta \tilde {z}, {\sigma}+\delta {\sigma})\,, \nonumber\\       
\end{eqnarray}

The evolution equations,upto first order, for these pertubations are
\begin{eqnarray}
     {\left( \begin{array}{cccc} \delta \dot {\tilde x} \\ \delta \dot {\tilde y} \\ \delta \dot {\tilde z} \\ \delta \dot \sigma \end{array} \right)    }=  {\bf{A}} {\left( \begin{array}{cccc} \delta {\tilde x} \\ \delta {\tilde y} \\ \delta {\tilde z}  \\ \delta \sigma \end{array} \right)}
\end{eqnarray}

where the matrix is 
\begin{eqnarray}
               {\bf A}=  {\left(
                 \begin{array}{cccc}
                      \frac{\partial f}{\partial \tilde x} & \frac{\partial f}{\partial \tilde y} & \frac{\partial f}{\partial \tilde z} & \frac{\partial f}{\partial \sigma} \\ 
                       \frac{\partial g}{\partial \tilde x} & \frac{\partial g}{\partial \tilde y} & \frac{\partial g}{\partial \tilde z} & \frac{\partial g}{\partial \sigma} \\ 
                        \frac{\partial h}{\partial \tilde x} & \frac{\partial h}{\partial \tilde y} & \frac{\partial h}{\partial \tilde z} & \frac{\partial h}{\partial \sigma}\\
                     \frac{\partial i}{\partial \tilde x} & \frac{\partial i}{\partial \tilde y} & \frac{\partial i}{\partial \tilde z} & \frac{\partial i}{\partial \sigma}
                  \end{array}
                 \right)}
\end{eqnarray}
is the Jacobian matrix and is evaluated at the fixed point $(\tilde{x}_c,\tilde{y}_c,\tilde{z}_c,\sigma_c)$ and hence each entry of $\bf{A}$ is a number. The solution of the system of equations can be found by diagonalizing the matrix $\bf{A}$. A non trivial solution exists only when the determinant $| \bf{A}- \lambda \bf{I}|$ is zero. Thus, solving this equation in $\lambda$ we would get all the eigen values of the system corresponding to each fixed points.

We have two cases: one with  positive kinetic term,  $sign(\rho_k)=+ve$ and other one with negative kinetic term,  $sign(\rho_k)=-ve.$.

\subsection{Closed Universe} \label{FPAClosed}
\subsubsection{Case I, with $sign\rho_k=+ve$}\label{ClosedCaseI}

In this case we study the fixed points for all possible values of parameters in an FRW closed universe. The fixed point $(0,0,0,0)$ is obtained for $w_k = w_m$ signifying all the dynamical variables $\tilde x$, $\tilde y$, $\tilde z$ and $\sigma$, going to zero at late times.  It is a nonhyperbolic fixed point as the eigen value of $\bf{A}$ for this is $(0,0,0,0)$. It's stability cannot be decided from our first order analysis of perturbations. From now onwards eigenvalues would mean eigenvalues of matrix $\bf{A}.$

\begin{figure} 
 \centering
 \includegraphics[width=0.60\textwidth]{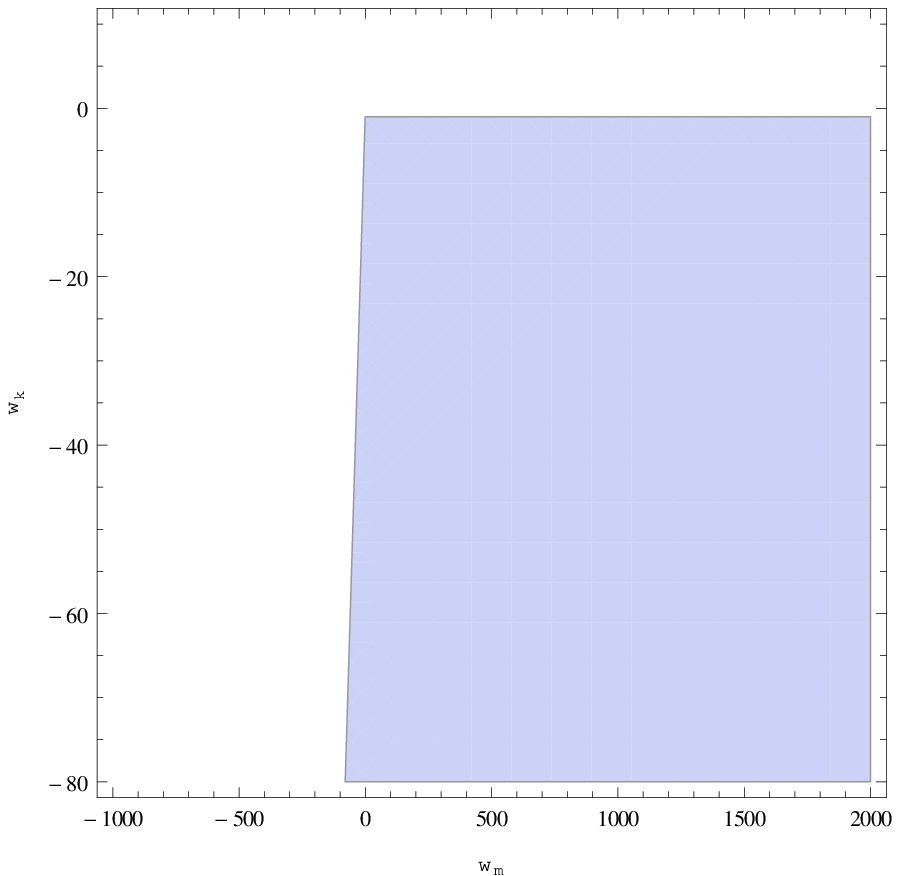}
 \caption{Allowed region of parameter space for the fixed point $(1,0,0,0)$ in closed universe}
 \label{ClosedParameterRegionI}
\end{figure}

 The second fixed point  $(1,0,0,0)$ denotes a late time kinetic dominated universe with other dynamical variables $\tilde y$, $\tilde z$ and $\sigma$ becoming zero. In this case eigenvalues are $(\frac{3(w_k+1)}{2},-1+\frac{3}{2}(1+w_k),\frac{3}{2}(-1+w_k+\frac{(1-w_k)(1+w_k)}{w_k}),3(w_k-w_m)).$  This is a stable fixed point for the region of parameter space shown in the Fig. [\ref{ClosedParameterRegionI}]. 

\begin{figure} 
 \centering
 \includegraphics[width=0.60\textwidth]{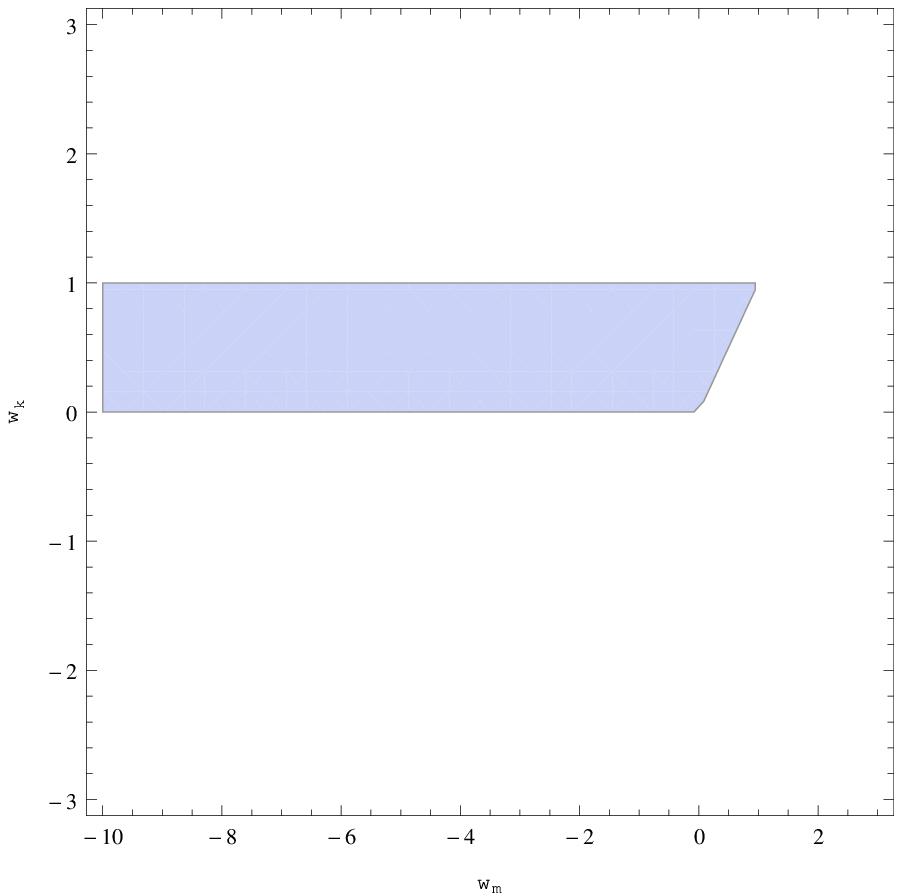}
 \caption{Allowed region of parameter space for the fixed point $(-1,0,0,0)$ in closed universe}
 \label{ClosedParameterRegionII}
\end{figure}

The next stable fixed point in this subsection is  $(-1,0,0,0)$  with eigenvalue $(1+\frac{3}{2}(-1-w_k),\frac{3}{2}(-1-w_k),\frac{3(-1-w_k)(-1+w_k+\frac{(1-w_k)(1+w_k)}{w_k})}{2(1+w_k)},\frac{3}{2}(-1-w_k)-\frac{3}{2}(1+w_k)+3(1+w_m))$ shows again a late time kinetic dominated phase but with a negative value of Hubble parameter $H$ signifying a contracting universe. This fixed point is found to be stable for the region of parameter space shown in Fig. [\ref{ClosedParameterRegionII}]. 
The point $(-1,0,0,0)$ may not be important from bouncing point of view, as we need the universe to transit to an expanding phase to be discussed in section \ref{BouncingScenario}. 

The remaining fixed points, in this section, being  $(0,0,\sqrt{3}\frac{\sqrt{-w_k+w_m}}{\sqrt{1+3w_m}},0)$  for $w_m>w_k$ and $w_m>-\frac{1}{3}$,  $(0,\frac{\sqrt{w_k-w_m}}{\sqrt{1+w_m}},0,0)$ with $w_k>w_m$ and $w_m>-1$, with eigen values \\$ (0,0,$\\$-\sqrt{\frac{3}{2}}\sqrt{w_k+3w_k^2-w_m-3w_kw_m},\sqrt{\frac{3}{2}}\sqrt{w_k+2w_k^2-w_m-3w_kw_m})$, $(0,0,-\frac{3}{2}\sqrt{\frac{w_k+w_k^2-w_m-w_kw_m}{2}},$\\$\frac{3}{\sqrt 2}\sqrt{w_k+w_k^2-w_m-w_kw_m})$ are also nonhyperbolic points. The stability of such fixed points goes beyond the linear stabilty analysis. 
All the fixed points and their stability conditions are noted in table\ref{tabl:Closedfixpt1}.

\begin{table}
 \centering
 \begin{tabular}{|l|l|l|r|}
  \hline
 Fixed Points $(x_c,y_c,z_c,\sigma_c)$ &              Stability Conditions      \\
  \hline 
   $(0,0,0,0)$ for $w_k=w_m$  &              Can't decide         \\
   $(1,0,0,0)$                &              Stable (see Fig.[\ref{ClosedParameterRegionI}])     \\
   $(-1,0,0,0)$               &              Stable (see Fig.[\ref{ClosedParameterRegionII}])     \\
   $(0,0,\sqrt{3}\frac{\sqrt{-w_k+w_m}}{\sqrt{1+3w_m}},0)$   with $w_m>w_k$ and $w_m>-\frac{1}{3}$  & Can't decide   \\
   $(0,\frac{\sqrt{w_k-w_m}}{\sqrt{1+w_m}},0,0)$             with $w_k>w_m$ and $w_m>-1$            & Can't decide   \\
  \hline
 \end{tabular}
 \caption{Stablity Analysis of fixed points for closed universe with $sign(\rho_k)=+ve$}
 \label{tabl:Closedfixpt1}
\end{table}

\subsubsection{Case II, $sign(\rho_k)=-ve$}\label{ClosedCaseII}

    In this section, we state the results of stability anlysis of our dynamical variables for the negative sign of kinetic energy density. The fixed points are found to be $(0, 0, 0, 0)$,$(0, 0, -\sqrt{3}\sqrt{\frac{w_k-w_m}{1+3w_m}}, 0)$ and $(0, \sqrt{\frac{-w_k+w_m}{1+w_m}}, 0, 0)$ with eigen values $(0, 0, 0, 0)$, $(0, 0, -\sqrt{\frac{3}{2}} \sqrt{-w_k-3w_k^2+w_m+3w_kw_m},$\\$
\sqrt{\frac{3}{2}}\sqrt{-w_k-3w_k^2+w_m+3w_k w_m})$ and $(‏0
, 0, -3 \frac{\sqrt{w_k-w_k^2+w_m+w_kw_m}}{\sqrt 2}, 3 \frac{\sqrt{w_k-w_k^2+w_m+w_kw_m}}{\sqrt 2})$ respectively. All these fixed points are nonhyperbolic and tabulated in table\ref{tabl:Closedfixpt2}.

\begin{table}
 \centering
 \begin{tabular}{|l|l|l|r|}
  \hline
 Fixed Points $(x_c, y_c, z_c, \sigma_c)$  & Stability Conditions      \\
  \hline 
   $(0, 0, 0, 0)$ for $w_k=w_m$                         &   Can't decide      \\
   $(0, 0, -\sqrt{3}\sqrt{\frac{w_k-w_m}{1+3w_m}}, 0)$         &  Can't decide     \\
   $(0, \sqrt{\frac{-w_k+w_m}{1+w_m}}, 0, 0)$                  &  Can't decide     \\
  \hline
 \end{tabular}
 \caption{Stablity Analysis of fixed points for closed universe with $sign(\rho_k)=-ve$}
 \label{tabl:Closedfixpt2}
\end{table}

\begin{figure}
 
 $
 \begin{array}{c c}
   \includegraphics[width=0.50\textwidth]{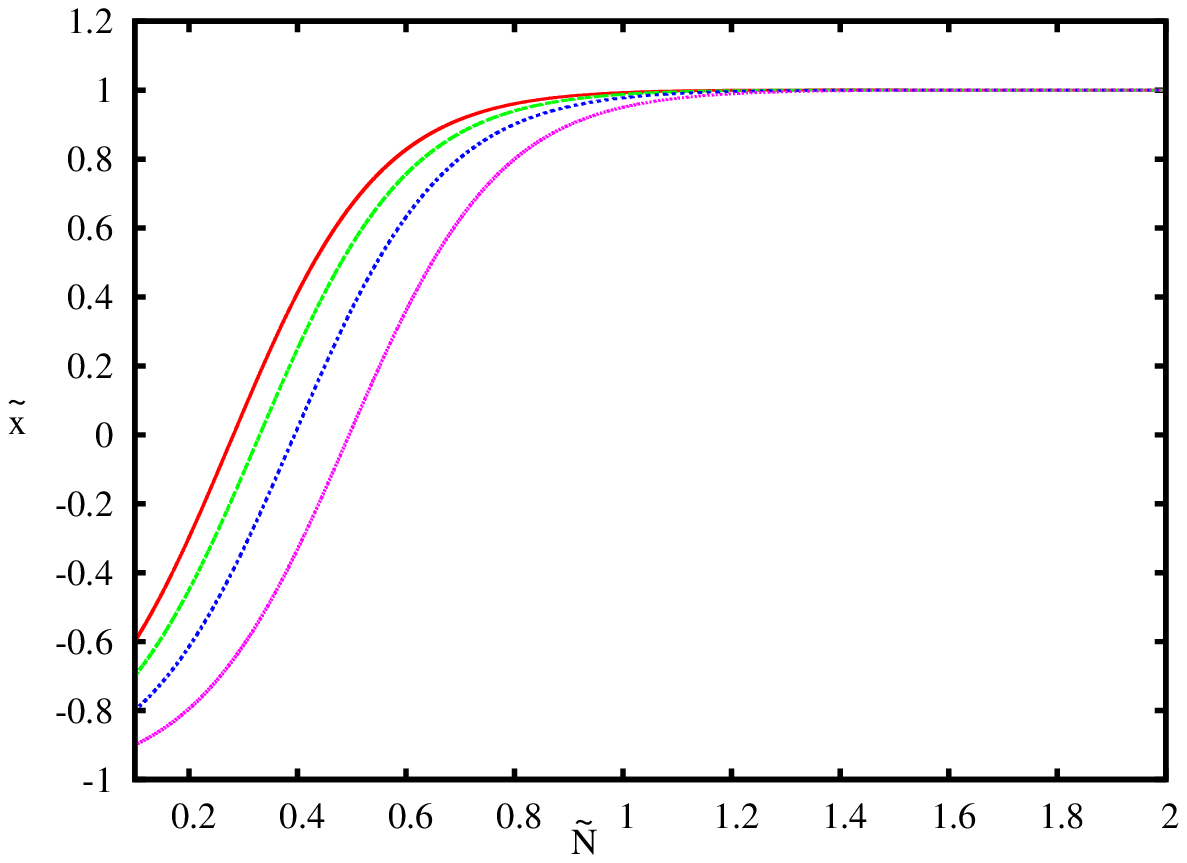} &
   \includegraphics[width=0.50\textwidth]{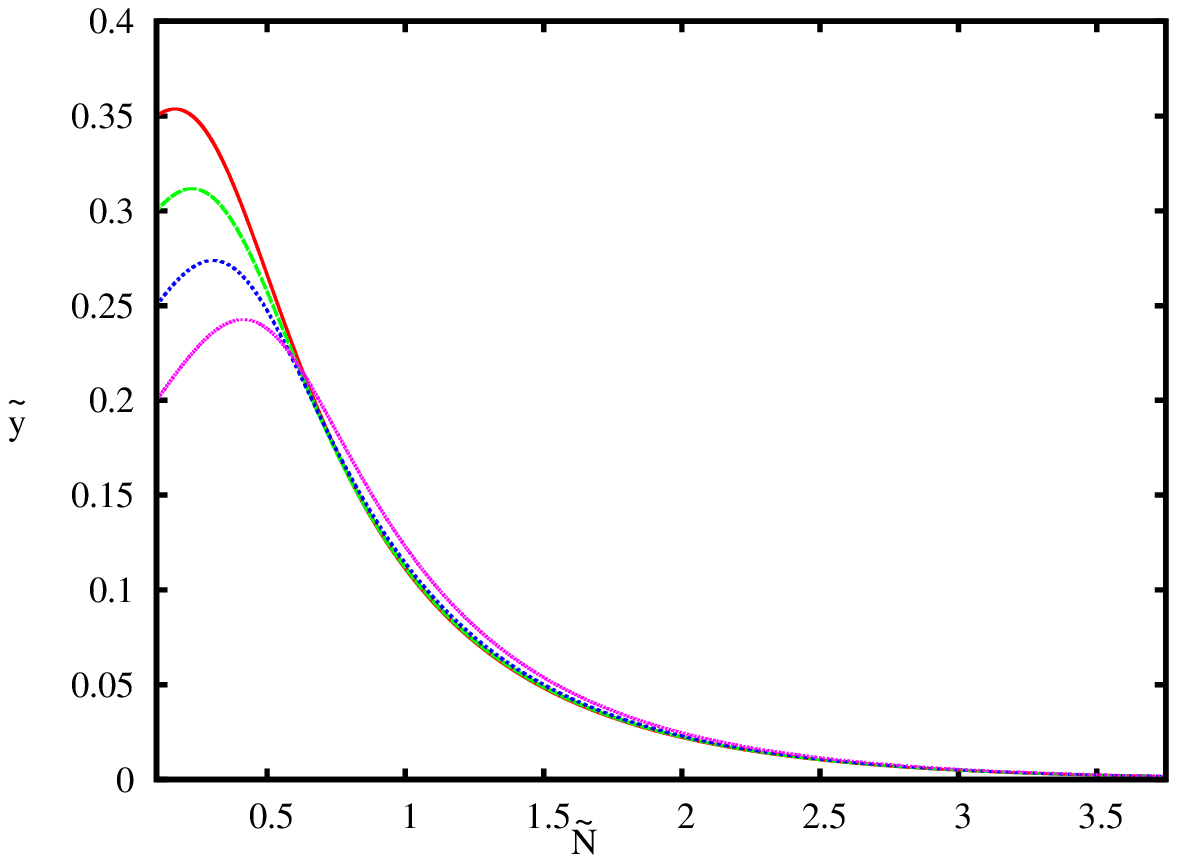} \\
   \includegraphics[width=0.50\textwidth]{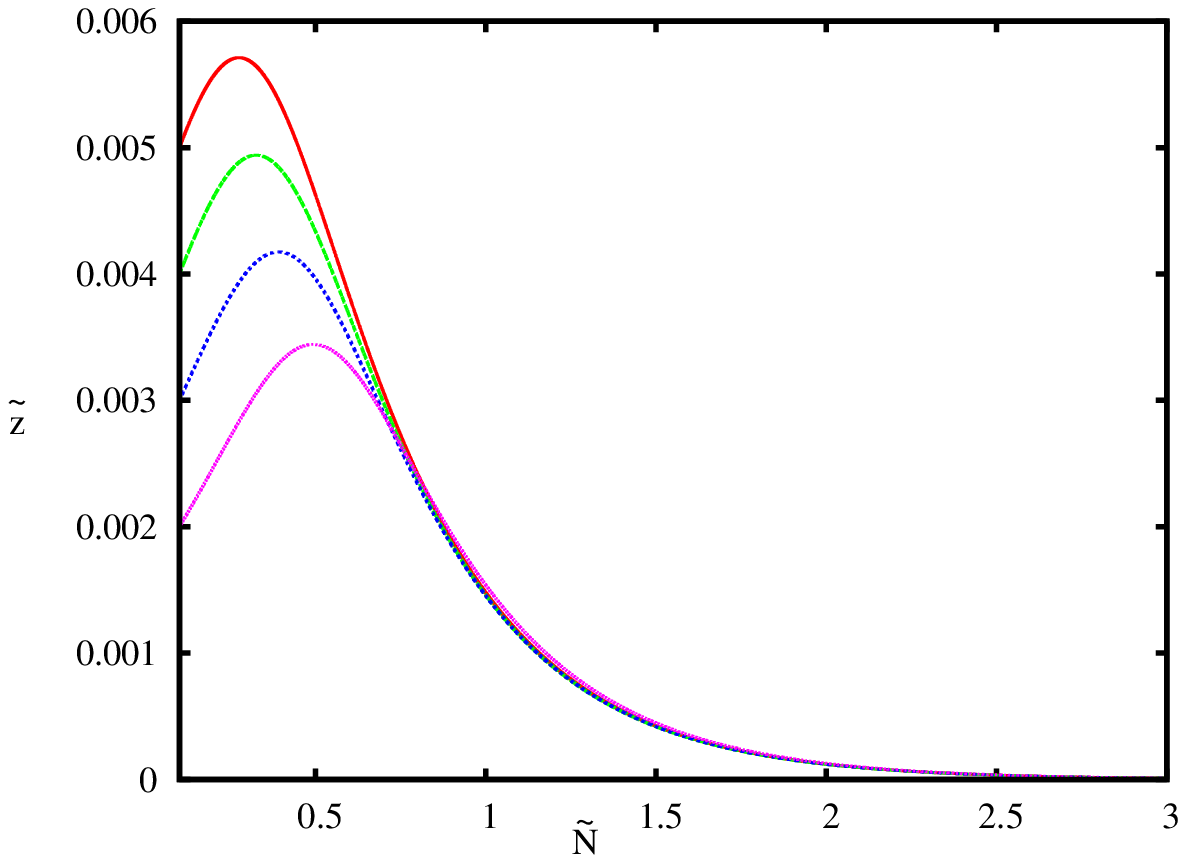} &
   \includegraphics[width=0.50\textwidth]{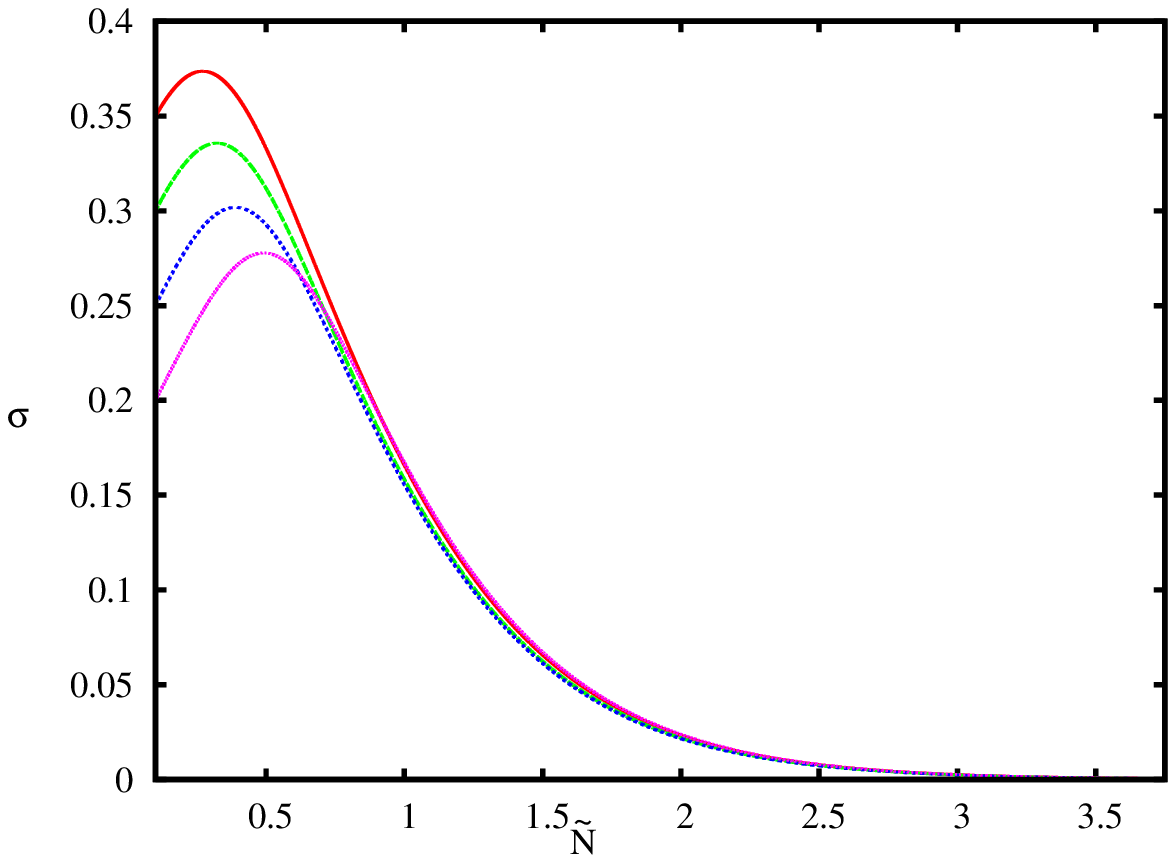}
 \end{array}
 $
  \caption{Evolution of the dynamical variables $\tilde x$ (\emph{top left}), $\tilde y$ (\emph{top right}), $\tilde z$ (\emph{bottom left}) and $\sigma$ (\emph{bottom right}) for the fixed point $(\tilde x_c, \tilde y_c, \tilde z_c, \sigma_c)=(1, 0, 0, 0)$  with the values of parameters $sign(\tilde z)=+ve$, $w_k=-2.0$, $w_m=1/3$ and $sign(\rho_k)=+ve$ for different initial conditions}
  \label{fig:ClosedFP1000}
\end{figure}

\subsection{Open Universe}\label{FPAOpen}
\subsubsection{Case I, with $sign\rho_k=+ve$}\label{OpenCaseI}

In this case we study the fixed points for all possible values of parameters in an FRW open universe. The fixed point $(0,0,0,0)$ is obtained for $w_k = w_m$ signifying all the dynamical variables $\tilde x$, $\tilde y$, $\tilde z$ and $\sigma$, going to zero at late times. It is a nonhyperbolic fixed point as the eigen value of $\bf{A}$ for this is $(0,0,0,0)$. It's stability cannot be decided from our first order analysis of perturbations.

\begin{figure} 
 \centering
 \includegraphics[width=0.60\textwidth]{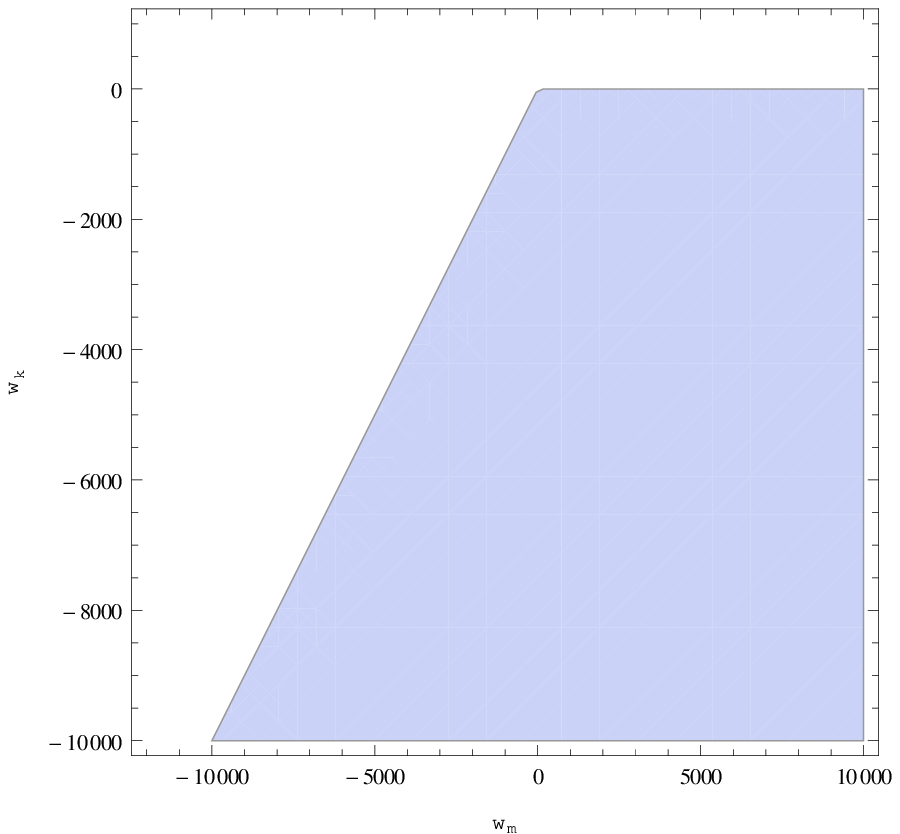}
 \caption{Allowed region of parameter space for the fixed point $(1,0,0,0)$ in open universe}
 \label{OpenParameterRegionI}
\end{figure}

 The second fixed point  $(1,0,0,0)$ denotes a late time kinetic dominated universe with other dynamical variables $\tilde y$, $\tilde z$ and $\sigma$ becoming zero. In this case eigenvalues are $(\frac{3(w_k+1)}{2},-1+\frac{3}{2}(1+w_k),\frac{3}{2}(-1+w_k+\frac{(1-w_k)(1+w_k)}{w_k}),3(w_k-w_m)).$  This is a stable fixed point for the region of parameter space shown in the Fig.\ref{OpenParameterRegionI}. 

\begin{figure} 
 \centering
 \includegraphics[width=0.60\textwidth]{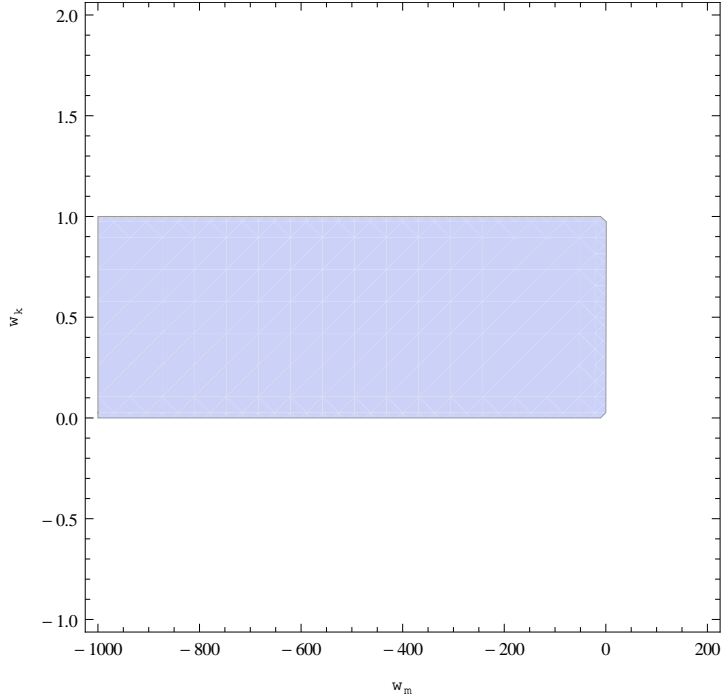}
 \caption{Allowed region of parameter space for the fixed point $(-1,0,0,0)$ in open universe}
 \label{OpenParameterRegionII}
\end{figure}

The next stable fixed point in this subsection is  $(-1,0,0,0)$  with eigenvalue $(1+\frac{3}{2}(-1-w_k),\frac{3}{2}(-1-w_k),\frac{3(-1-w_k)(-1+w_k+\frac{(1-w_k)(1+w_k)}{w_k})}{2(1+w_k)},\frac{3}{2}(-1+w_k)-\frac{3}{2}(1+w_k)+3(1+w_m))$ shows again a late time kinetic dominated phase but with a negative value of Hubble parameter $H$ signifying a contracting universe. This fixed point is found to be stable for the region of parameter space shown in Fig.[\ref{OpenParameterRegionII}]. 
The point $(-1,0,0,0)$ may not be important for bouncing point of view, as again, we need the universe to transit to an expanding phase to be discussed in section \ref{BouncingScenario}. 

The remaining two fixed points being  $(0,0,-\frac{\sqrt 3\sqrt{w_k-w_m}}{\sqrt{1+3w_m}},0)$ with $w_k>w_m$ and $w_m>-\frac{1}{3}$ and $(0,\frac{\sqrt{w_k-w_m}}{\sqrt{1+w_m}},0,0)$  with $w_k>w_m$ and $w_m>-1$  with eigen values\\ $ (0,0,-\sqrt{\frac{3}{2}}\sqrt{w_k+3w_k^2-w_m-3w_kw_m},\sqrt{\frac{3}{2}}\sqrt{w_k+3w_k^2-w_m-3w_kw_m})$ and\\ $(0,0,-\frac{3}{\sqrt 2}\sqrt{w_k+w_k^2-w_m-w_kw_m},\frac{3}{\sqrt 2}\sqrt{w_k+w_k^2-w_m-w_kw_m})$ are also nonhyperbolic points. The stability of such fixed points goes beyond the linear stabilty analysis.        
All the fixed points and their stability are noted in table \ref{tabl:Openfixpt1}.

\begin{table}
 \centering
 \begin{tabular}{|l|l|l|r|}
  \hline
 Fixed Points $(x_c,y_c,z_c,\sigma_c)$ &              Stability Conditions      \\
  \hline 
   $(0,0,0,0)$ for $w_k=w_m$  &              Can't decide        \\
   $(1,0,0,0)$                &              Stable (see Fig.[\ref{OpenParameterRegionI}])     \\
   $(-1,0,0,0)$               &              Stable (see Fig.[\ref{OpenParameterRegionII}])     \\
   $(0,0,-\frac{\sqrt 3\sqrt{w_k-w_m}}{\sqrt{1+3w_m}},0)$  with $w_k>w_m$ and $w_m>-\frac{1}{3}$  & Can't decide \\
   $(0,\frac{\sqrt{w_k-w_m}}{\sqrt{1+w_m}},0,0)$  with $w_k>w_m$ and $w_m>-1$             & Can't decide  \\
  \hline
 \end{tabular}
 \caption{Stablity Analysis of fixed points for open universe with $sign(\rho_k)=+ve$}
 \label{tabl:Openfixpt1}
\end{table}

\subsubsection{Case II, $sign(\rho_k)=-ve$}\label{OpenCaseII}

    In this section, we state the results of stability anlysis of our dynamical variables for the negative sign of kinetic energy density. The fixed points are found to be $(0, 0, 0, 0)$,$(0, 0, -\sqrt {\frac{3(-w_k+w_m)}{1+3w_m}}, 0)$  with $w_m>w_k$ and $w_m>-\frac{1}{3}$, and $(0, \sqrt{\frac{-w_k+w_m}{1+w_m}}, 0, 0)$ with $w_m>w_k$ and $w_m>-1$ with eigen values $(0, 0, 0, 0)$, $(0, 0,-\sqrt{\frac{3}{2}}\sqrt{-w_k-3w_k^2+w_m+3w_kw_m} ,\sqrt{\frac{3}{2}}\sqrt{-w_k-3w_k^2+w_m+3w_kw_m})$ and $(‏0,0,-\frac{3\sqrt{-w_k-w_k^2+w_m+w_kw_m}}{\sqrt 2},\frac{3\sqrt{-w_k-w_k^2+w_m+w_kw_m}}{\sqrt 2})$ respectively.  All these fixed points are nonhyperbolic and tabulated in table. \ref{tabl:Openfixpt2}.

\begin{table}
 \centering
 \begin{tabular}{|l|l|l|r|}
  \hline
 Fixed Points $(x_c, y_c, z_c, \sigma_c)$  & Stability Conditions      \\
  \hline 
   $(0, 0, 0, 0)$ for $w_k=w_m$                                                                         &   Can't decide       \\
   $(0, 0, -\sqrt {\frac{3(-w_k+w_m)}{1+3w_m}}, 0)$  with $w_m>w_k$ and $w_m>-\frac{1}{3}$              &   Can't decide     \\
   $(0, \sqrt{\frac{-w_k+w_m}{1+w_m}}, 0, 0)$ with $w_m>w_k$ and $w_m>-1$                               &   Can't decide       \\
  \hline
 \end{tabular}
 \caption{Stablity Analysis of fixed points for open universe with $sign(\rho_k)=-ve$}
 \label{tabl:Openfixpt2}
\end{table}

\begin{figure}
 
 $
 \begin{array}{c c}
   \includegraphics[width=0.50\textwidth]{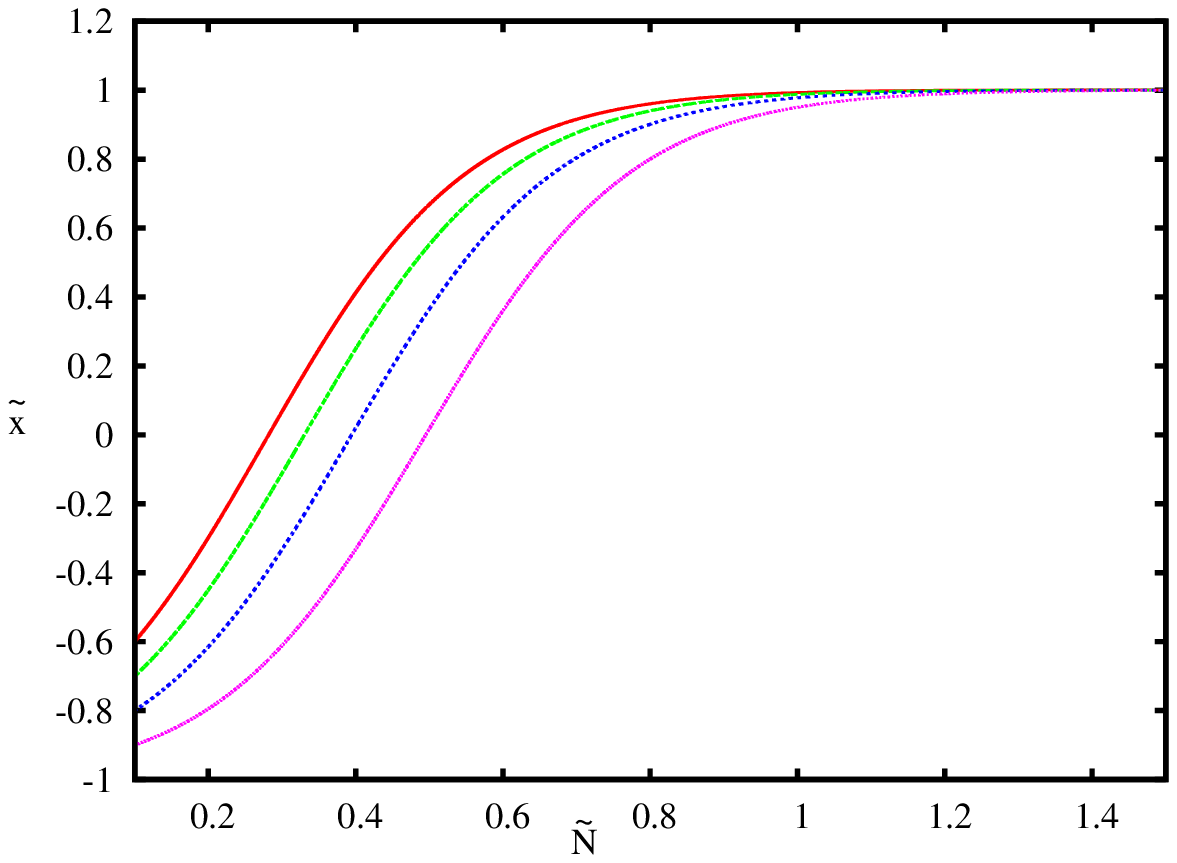} &
   \includegraphics[width=0.50\textwidth]{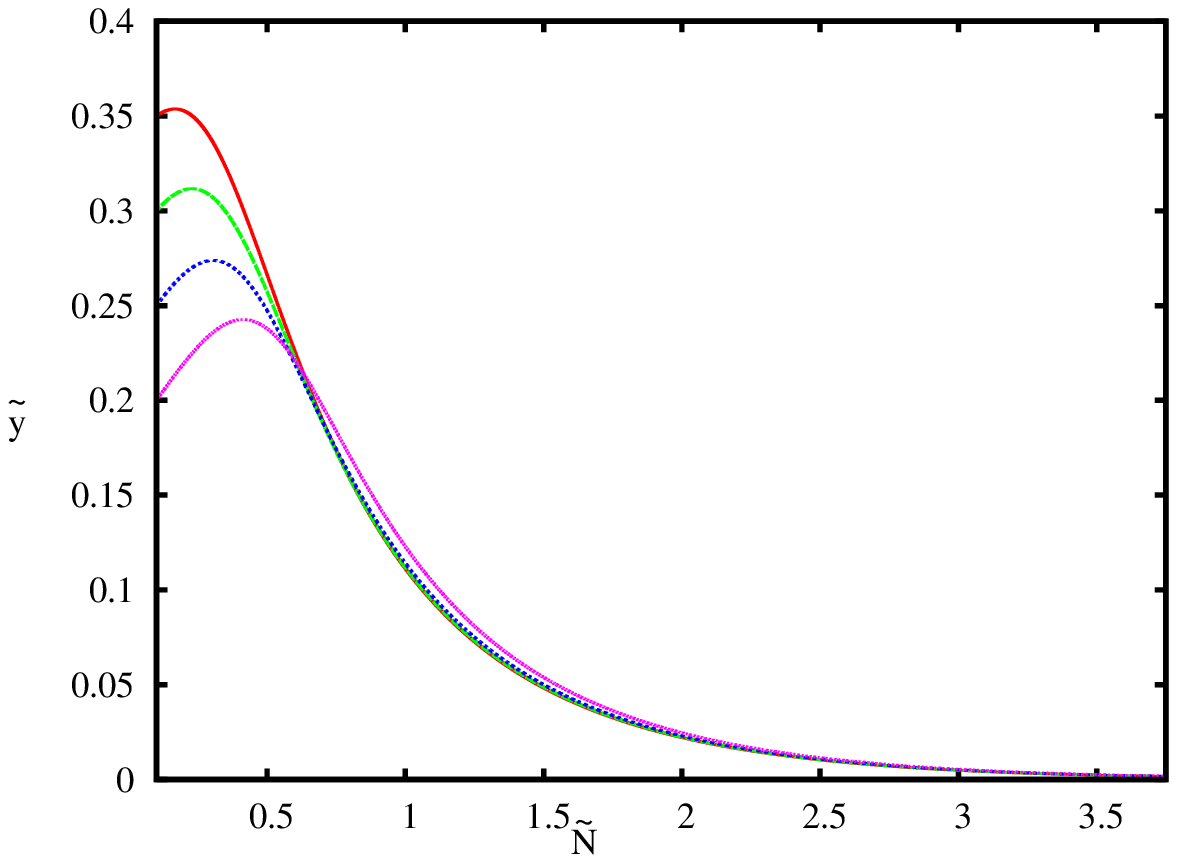} \\
   \includegraphics[width=0.50\textwidth]{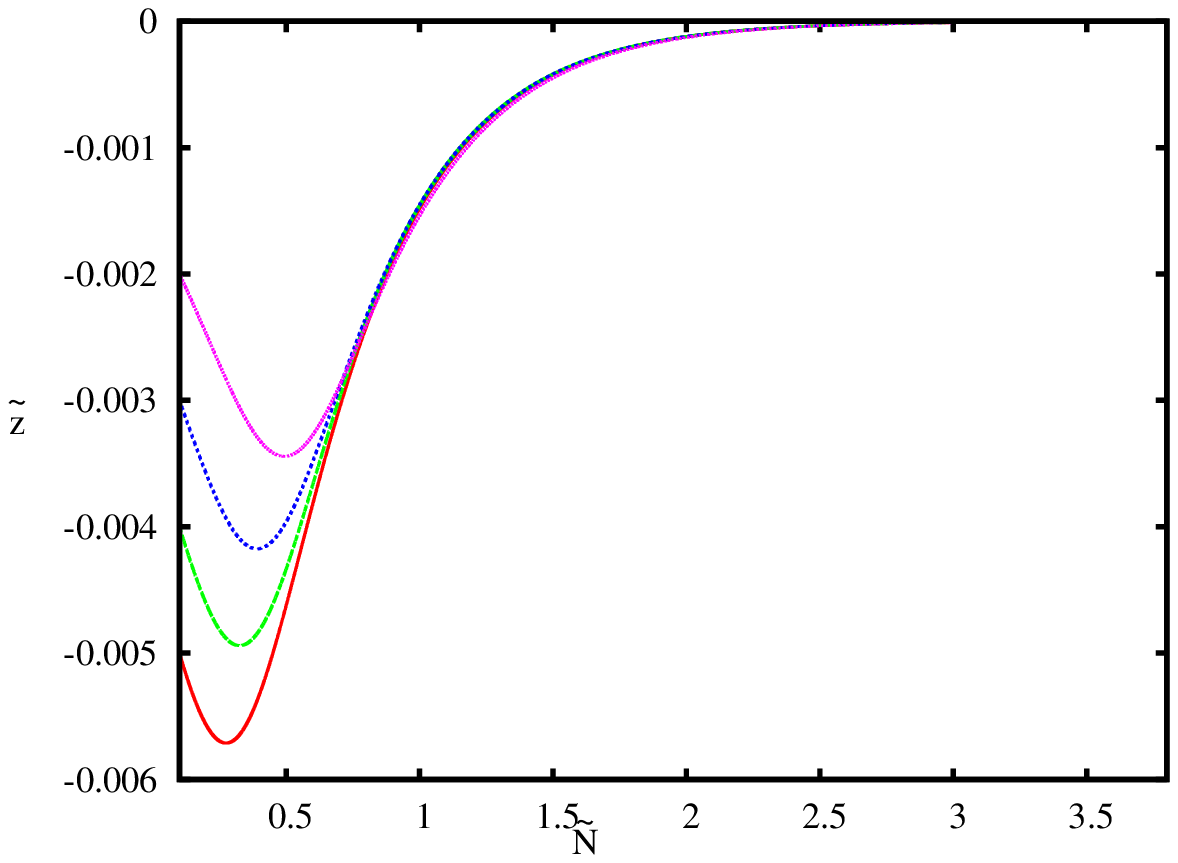} &
   \includegraphics[width=0.50\textwidth]{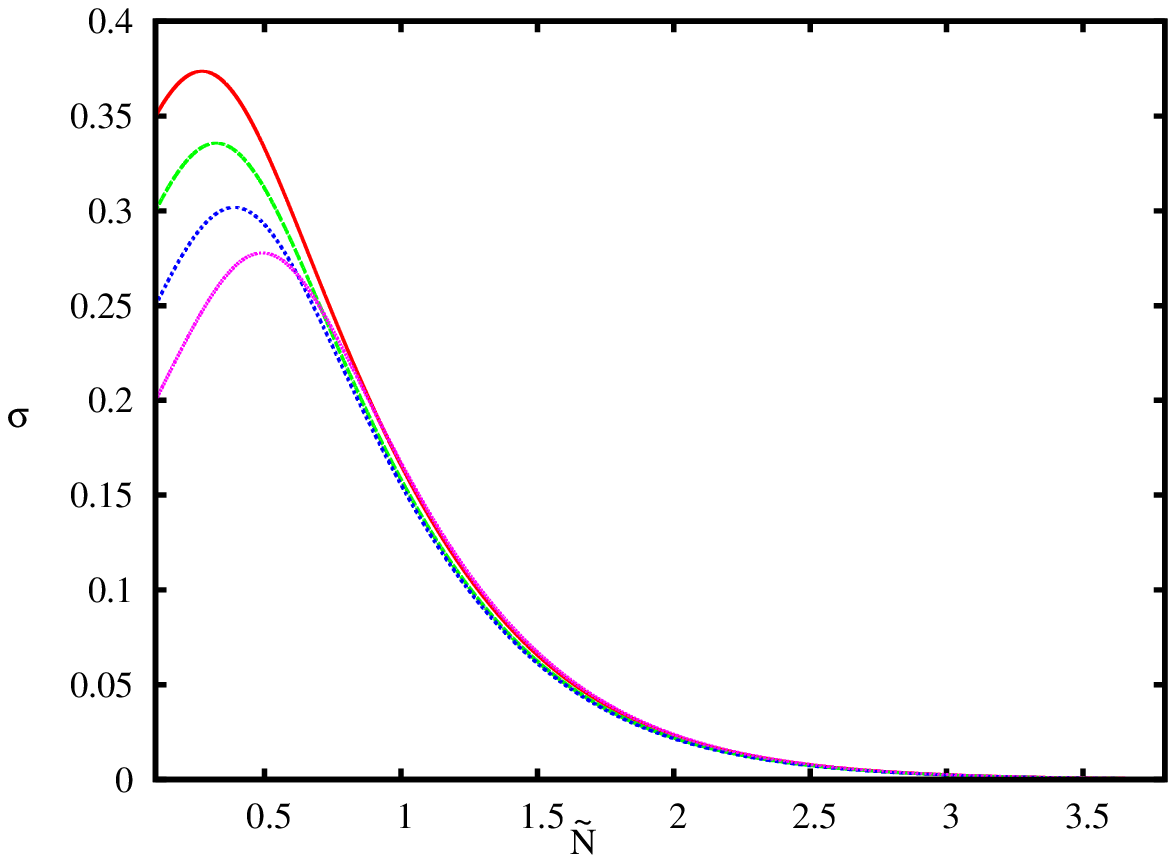}
 \end{array}
 $
  \caption{Evolution of the dynamical variables $\tilde x$ (\emph{top left}), $\tilde y$ (\emph{top right}), $\tilde z$ (\emph{bottom left}) and $\sigma$ (\emph{bottom right}) for the fixed point $(\tilde x_c, \tilde y_c, \tilde z_c, \sigma_c)=(1, 0, 0, 0)$  with the values of parameters $sign(\tilde z)=-ve$, $w_k=-2.0$, $w_m=1/3$ and $sign(\rho_k)=+ve$ for different initial conditions}
  \label{fig:OpenFP1000}
\end{figure}

\section{Bouncing Scenario}\label{BouncingScenario}

Now we obtain the conditions for non singular bounce to occur and also show the evolution of dynamical variables numerically. A nonsingular bounce is attained whenever the universe passes from a contracting phase to an expanding phase through a minimum value of the avearge scale factor $a(t),$ but not zero. Mathematically, it satisfies
\begin{equation}
(H)_b\equiv \frac{1}{a_b(t)}\left ( \frac{\mathrm{d} a(t)}{\mathrm{d} t} \right )_b=0 ,
\end{equation}
where subscript $b$ denotes value of the variable at the bounce, and 
\begin{equation}
\left ( \frac{\mathrm{d}^2 a(t)}{\mathrm{d} t^2} \right )_b>0
\end{equation}
for minimum to occur. This implies 
\begin{equation}
\left ( \frac{\mathrm{d} H}{\mathrm{d} t} \right )_b=\left ( \frac{\ddot{a}}{a} \right )_b-\left ( \frac{\dot{a}}{a} \right )^2_b>0
\end{equation}

Now, writing the above conditions in terms of dynamical variables for bouncing, we get $\tilde{x}_b=0$  and $\left ( \frac{\mathrm{d} \tilde{x}}{\mathrm{d} \tilde{N}} \right )_b>0$ which translate to the following equation 
\begin{equation}
\left ( \frac{\mathrm{d} \tilde{x}}{\mathrm{d} \tilde{N}} \right )_b
=-\frac{3}{2}\left [ (w_k-w_m)(sign\rho_k)+(1+w_m)(-\tilde{y}|\tilde{y}|)+\frac{(1+3w_m)}{3}\tilde{z}|\tilde{z}| \right ]>0. 
\label{bcondition1}
\end{equation}
This implies
\begin{equation}
\left({\tilde{y}\tilde{|y|}}{(1+w_m)}-{\tilde{z}|\tilde z|}{(1+3w_m)}\right)_b > 1\times sign(\rho_k)({w_k-w_m}).
\label{bcondition2}
\end{equation} 

At the bounce we then obtain the constraint equation among dynamical variable as
\begin{equation}
\left ( \tilde{x}^2 -\tilde{y}\tilde{|y|}+\tilde{z}|\tilde{z}|-\tilde{\Omega}_m\right )_b = -\tilde{y}\tilde{|y|}+\tilde{z}|\tilde{z}|-\tilde{\Omega}_m=1\times sign(\rho_k).
\label{bcondition3}
\end{equation} 

\begin{figure}
 \includegraphics[width=0.48\textwidth]{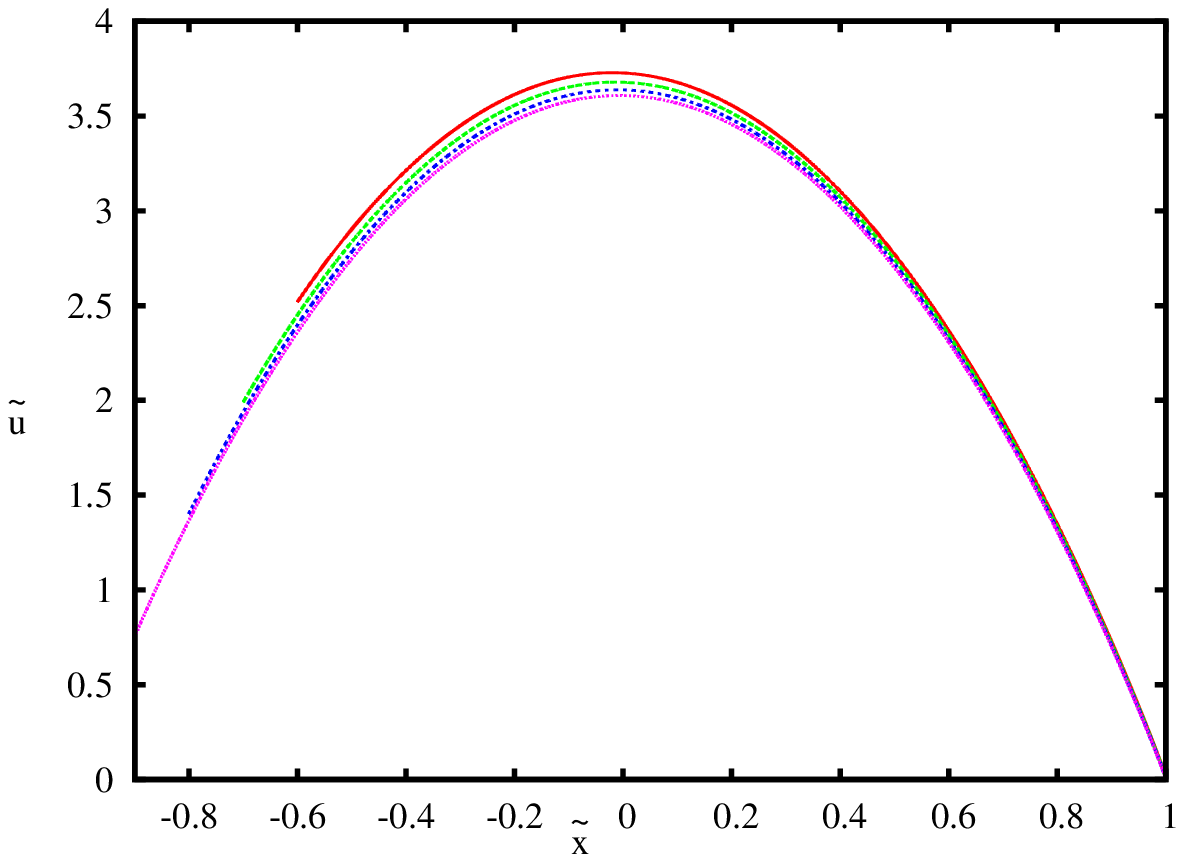} 
 \includegraphics[width=0.48\textwidth]{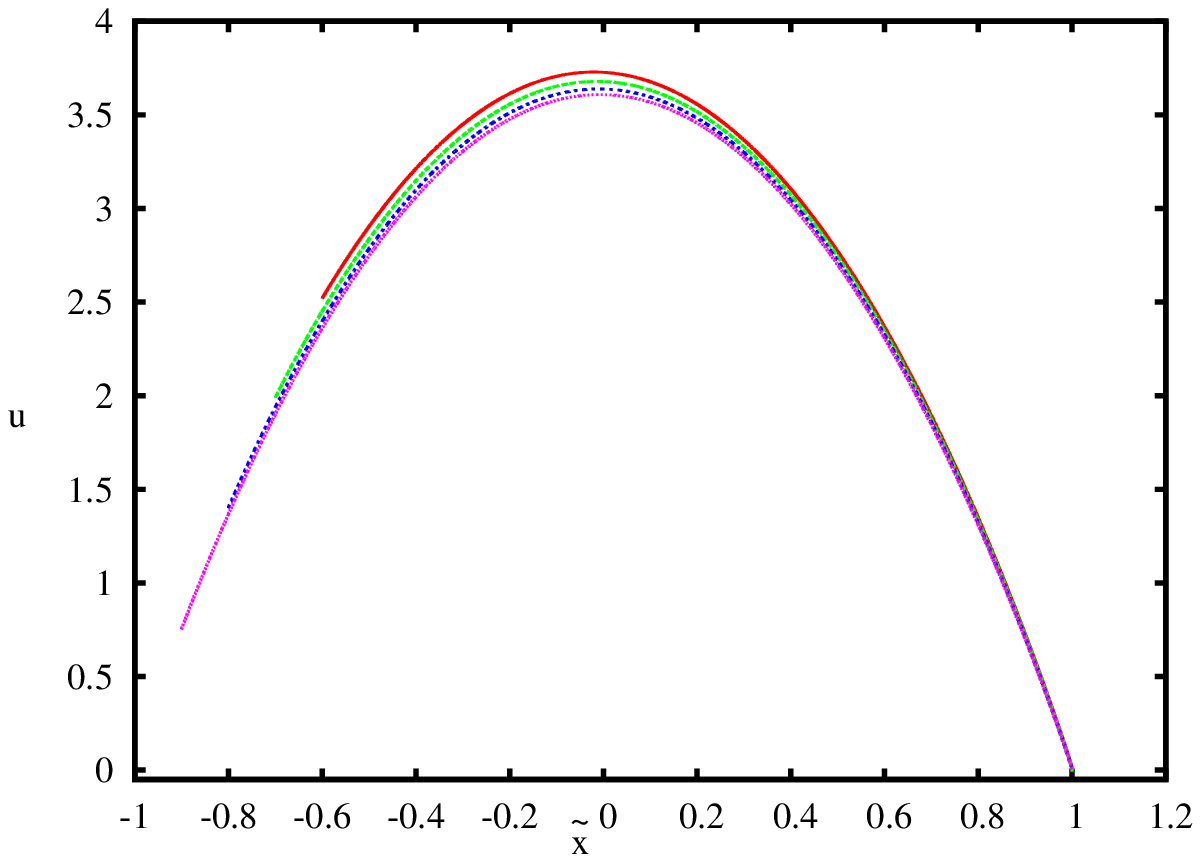}
 \caption{$u\equiv \frac{\mathrm d \tilde x}{\mathrm d \tilde N}$ vs $\tilde x$ for closed  (\emph{left}) and for open (\emph{right}) with $w_k=-2.0$, $w_m=1/3$ and $sign(\rho_k)=+ve$}
 \label{Slope}
\end{figure}

           Now, for different negative initial conditions of $\tilde x$ (contracting phase), Fig. [\ref{fig:ClosedFP1000}] (\emph{top left})   and [\ref{fig:OpenFP1000}] (\emph{top left})for closed and open universe  respectively, show its transition to positive values (expanding phase) crossing zero (bounce). The bouncing is guarenteed by the positivity of the slope of $\tilde x$ as shown in Fig. [\ref{Slope}] (left plot  for closed and right plot for open). Thus, top left of Fig. [\ref{fig:ClosedFP1000}] and Fig. [\ref{fig:OpenFP1000}] and Fig. [\ref{Slope}], together, do indeed represent stable bouncing scenario in FRW closed and open universe. This is obtained by setting the values of equation of state parameters $w_k=-2$ $(\eta=1/4), $ $w_m=1/3$ and $sign(\rho_k)=+ve$ and $sign(y)=+ve$. The evolution of other dynamical variables can be seen in Fig. [\ref{fig:ClosedFP1000}] and [\ref{fig:OpenFP1000}], which show their asymptotic evolution to the respective fixed points for the same choice of parameters. 

           It can be seen that the fixed point $(\tilde x_c, \tilde y_c, \tilde z_c, \sigma_c)=(1,0,0,0)$ does give rise to a stable bouncing universe as it satisfies Eqs.(\ref{bcondition2}) and \ref{bcondition3}) for open ($sign(\tilde z)$=-ve) and closed ($sign(\tilde z)$=+ve) universe. From this analysis, we conclude that finally after the bounce our universe at late times is driven by kinetic energy density in both the cases. 
The other fixed point $(-1,0,0,0)$, though stable, can not give rise to a bouncing scenario as it ends up with a negative value of Hubble parameter, $H$, signifying a late time contracting phase. 

\begin{figure}
 \includegraphics[width=0.48\textwidth]{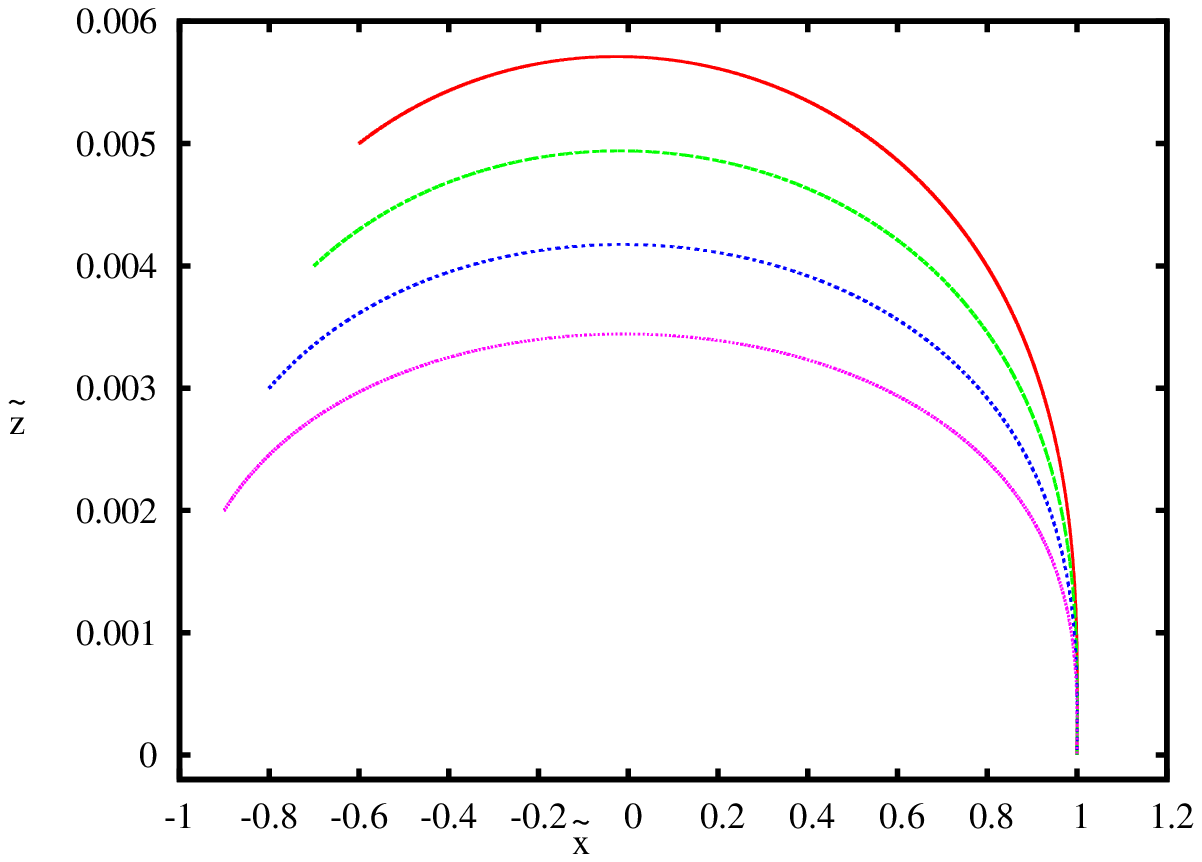} 
 \includegraphics[width=0.48\textwidth]{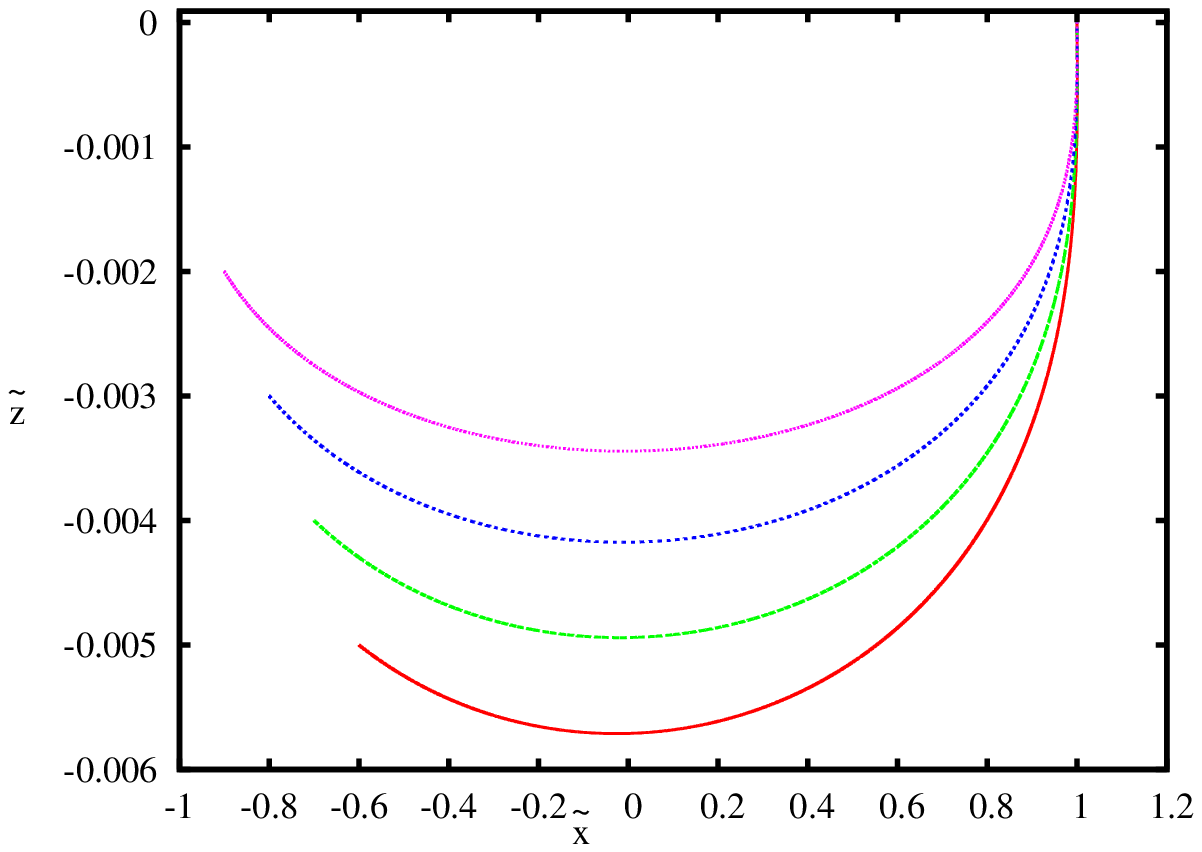}
 \caption{$\tilde z$ vs $\tilde x$ for closed (\emph{left})  and for open  (\emph{right}) with $w_k=-2.0$, $w_m=1/3$ and $sign(\rho_k)=+ve$}
 \label{Curvature}
\end{figure}

          Also, we show the behaviour of curvature parameter, $\tilde z,$ in this nonsingular bouncing set up. The curvature parameter increases initially in the contracting phase reaching an extremum at the bounce and then decreases to zero in the expanding phase as shwon in Fig. [\ref{Curvature}] for both open and closed universe. Thus, the curvature parameter remains finite at the bounce as expected in a nonsingular bouncing scenario and at late time universe becomes flat irrespective of whether we start initially with closed or open.
This may be useful for building realistic models.

\begin{figure}
 \includegraphics[width=0.48\textwidth]{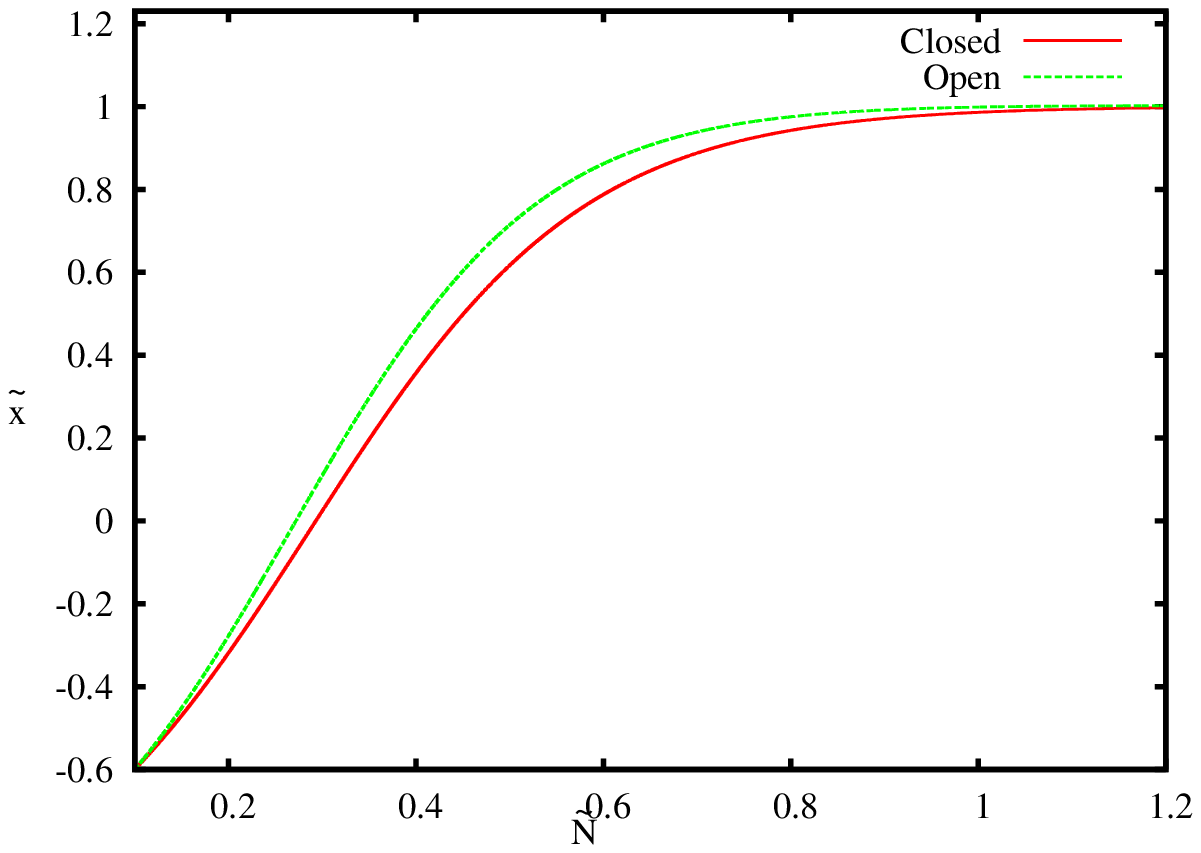} 
 \includegraphics[width=0.48\textwidth]{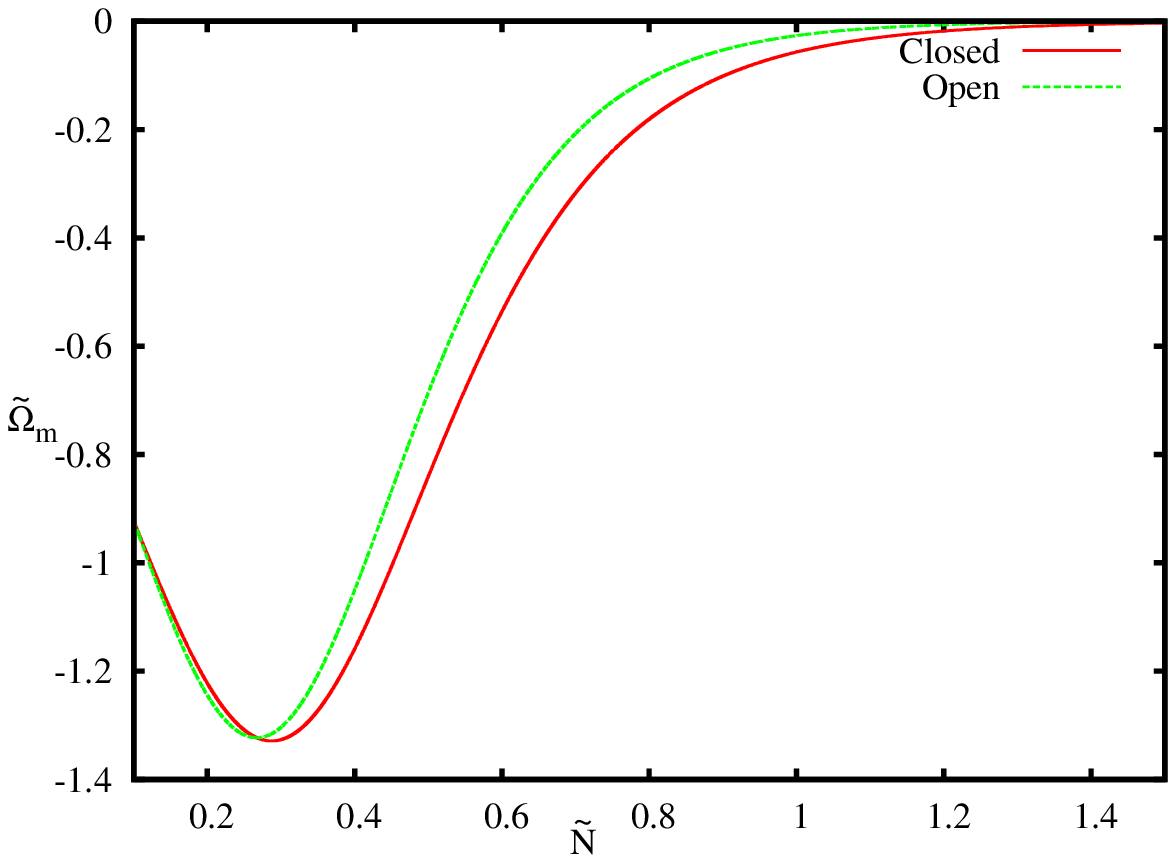}
 \caption{Comparision between closed and open for $\tilde x$ vs $\tilde N$ (\emph{left}) and $\tilde {\Omega}_m$ vs $\tilde N$ (\emph{right}) with $w_k=-2.0$, $w_m=1/3$ and $sign(\rho_k)=+ve$}
 \label{fig:CompareBouncing}
\end{figure}

The comparision between bouncing solutions for open and closed is done in Fig. [\ref{fig:CompareBouncing}]. It has been found that bouncing occurs earlier in the case of open than in the closed universe as shown in left hand side of Fig. [\ref{fig:CompareBouncing}] for the same set of initial conditions and parameters. Also, it is noted that, though, the solutions differ appreciably near the bounce, they approach to the same value at late time owing to zero value of the curvature parameter. The nonsingular bounce happens only for negative values of $\tilde{\Omega}_m$ with our choice of parameters as shown in Fig. [\ref{fig:CompareBouncing}] (\emph{right}) for both open and closed universe.

\begin{figure}
 
 $
 \begin{array}{c c}
   \includegraphics[width=0.50\textwidth]{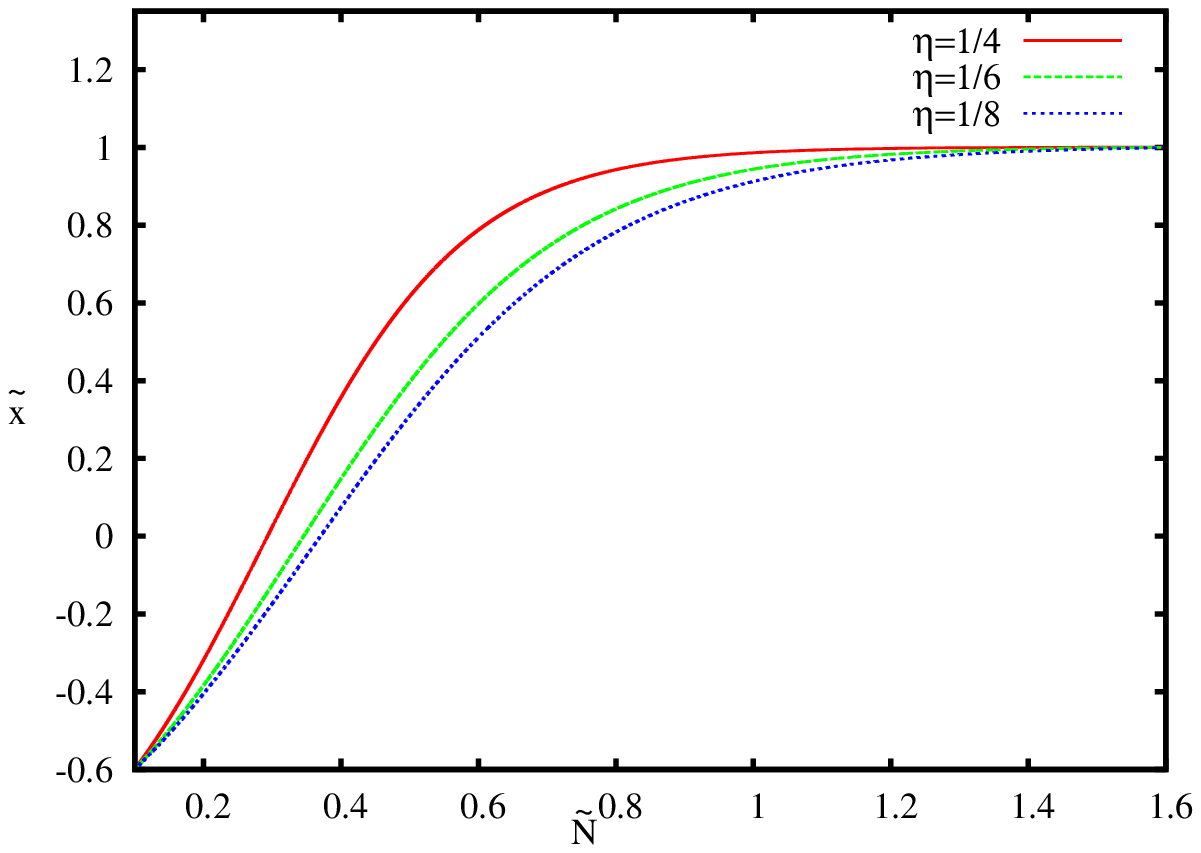} &
   \includegraphics[width=0.50\textwidth]{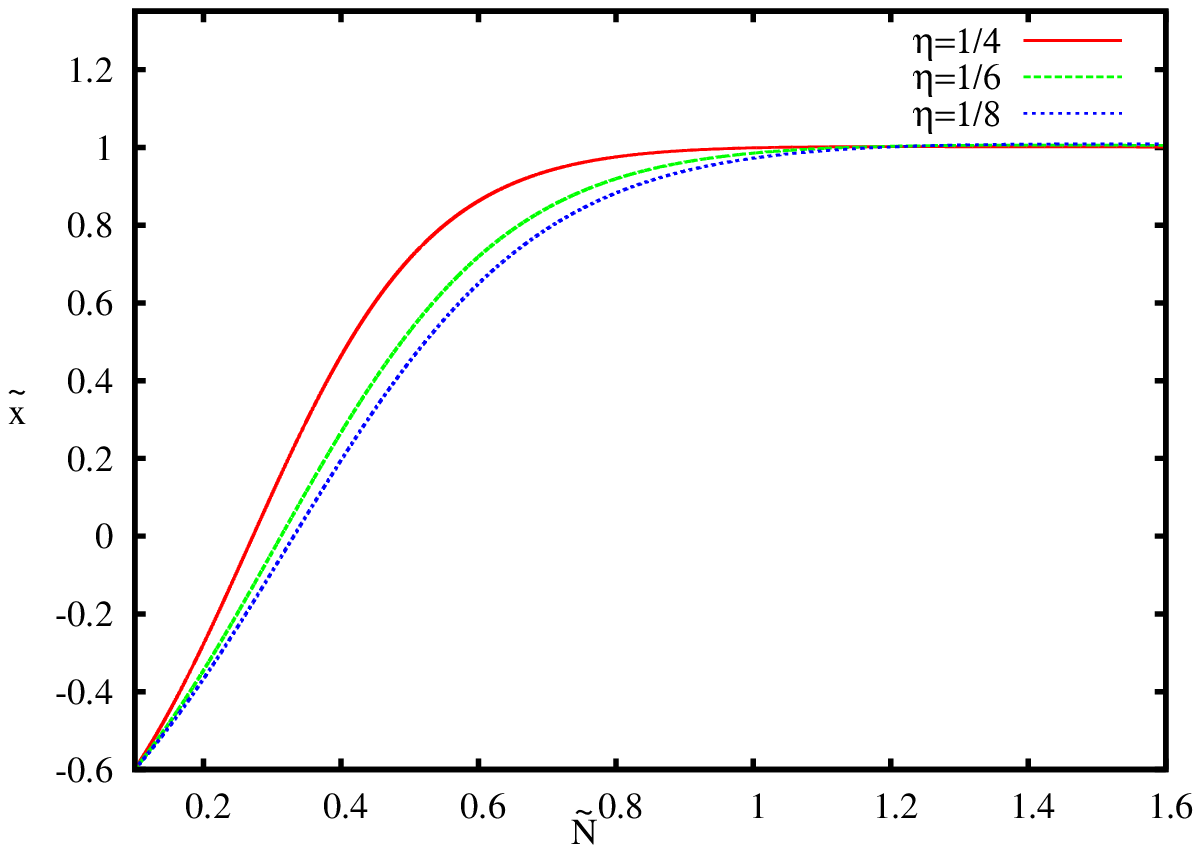} \\
   \includegraphics[width=0.50\textwidth]{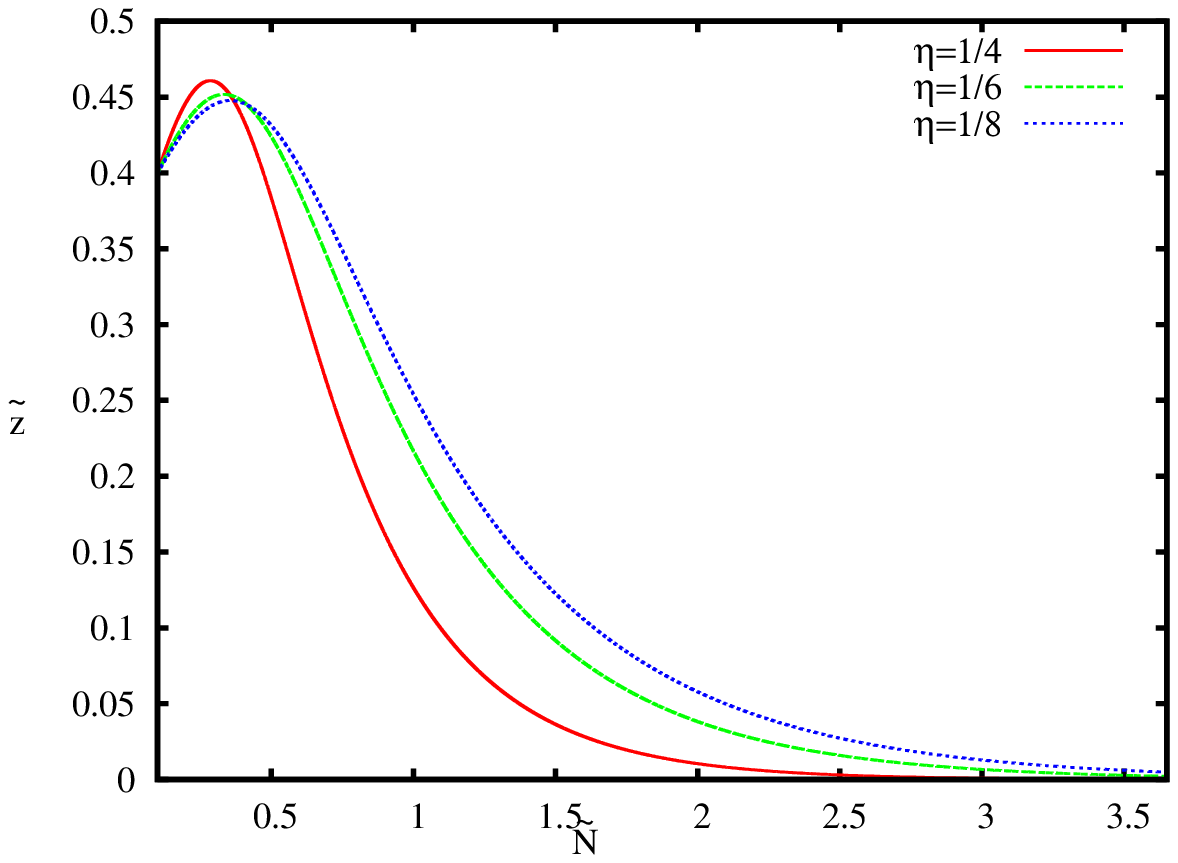} &
   \includegraphics[width=0.50\textwidth]{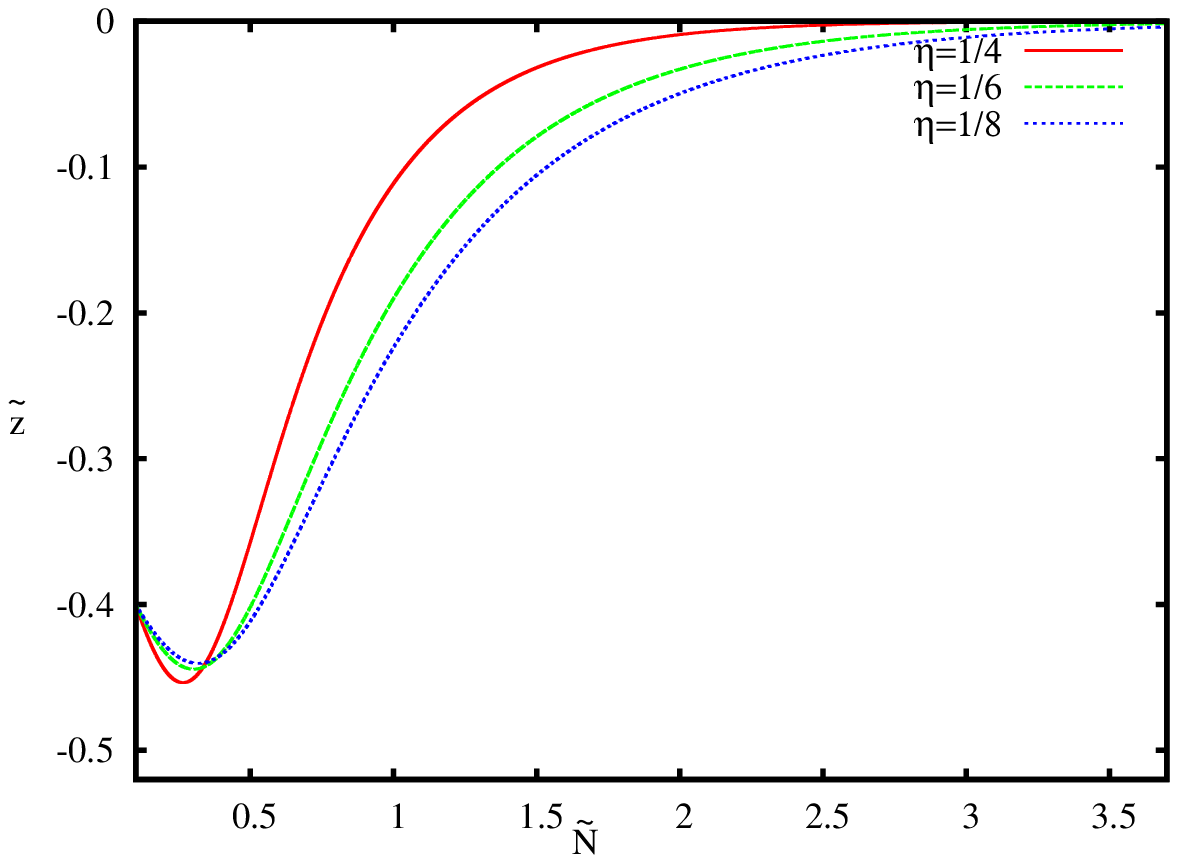}
 \end{array}
 $
  \caption{Evolution $\tilde x$ for closed (\emph{top left})  and open (\emph{top right}); $\tilde z$ for closed (\emph{bottom left}) and open (\emph{bottom right}) with values $\eta=1/4,$$1/6,$ and $1/8$, $w_m=1/3$ and $sign(\rho_k)=+ve$}
  \label{VaryEta}
\end{figure}

         Finally we show the effect of different values of $\eta$ on the behaviour of bouncing solutions in Fig. [\ref{VaryEta}] for both closed and open universe.  All the plots are generated for the same set of initial conditions and the same set of parameters $w_m=1/3$, $sign(\rho_k)=+ve$ but with three different values of parameters $\eta=$$1/4$, $1/6$ and $1/8$. It has been observed that the value of $\eta$ has a direct impact on the occurence of bouncing point. Indeed, the position of bouncing point is delayed as we decrease the value of $\eta$ for both closed and open universe as shown in Fig. [\ref{VaryEta}] (\emph{top left} and \emph{top right}). The bottom left and bottom right of Fig. [\ref{VaryEta}] indicate the effect of $\eta$ on the curvature parameter for both closed and open cases respectively.
It has been found that the magnitude of maximum value of $\tilde z,$ at the bounce, decreases as we decreases the value of $\eta.$

\section{Conclusion}\label{Conclusion}

A cosmological scenario with a noncanonical scalar field and matter is explored in this work. Using dynamical equations for a set of dimensionless dynamical variables, we find all fixed point for the two cases with positive and negative kinetic energy density term in FRW closed and open universe. Allowed region of parameter spaces for stability of fixed points are shown for both cases. The necessary and sufficient conditions for a nonsingular bounce are obtained in terms of the dynamical variables.
Thus, stable bouncing solutions are obtained satisfying nonsingular bouncing conditions and stability criteria.
This is achieved for the negative energy density of matter, $\tilde{\Omega}_m,$ with equation of state parameter $w_m=1/3$ in both closed and open universe. In addition to this, the finitude of curvature parameter at the bounce is obtained as expected in a nonsingular bouncing scenario and universe becomes flat at late time irrespective of whether we start with a closed or open one. Finally, the effect of the parameter $\eta$ on the behaviour of bouncing solution is noted. It is seen that the point of occurence of bounce is delayed  as we decrease the value of $\eta$ and the magnitude of value of curvature parameter at the bounce decreases with $\eta$ for both open and closed universe.
 
We restrict our analysis to a positive sign of potential. It is straightforward to extend our analysis for a negative potential by changing the parameter $sign(y)$ to $-1$.

\end{document}